\PassOptionsToPackage{unicode}{hyperref}
\PassOptionsToPackage{hyphens}{url}
\PassOptionsToPackage{dvipsnames,svgnames,x11names}{xcolor}
\documentclass[
  a4paperpaper,
  english,
  abstract=true]{scrartcl}
\usepackage{xcolor}
\usepackage{amsmath,amssymb}
\usepackage{iftex}
\ifPDFTeX
  \usepackage[T1]{fontenc}
  \usepackage[utf8]{inputenc}
  \usepackage{textcomp} 
\else 
  \usepackage{unicode-math} 
  \defaultfontfeatures{Scale=MatchLowercase}
  \defaultfontfeatures[\rmfamily]{Ligatures=TeX,Scale=1}
\fi
\usepackage[]{libertinus}
\ifPDFTeX\else
\fi
\IfFileExists{upquote.sty}{\usepackage{upquote}}{}
\IfFileExists{microtype.sty}{
  \usepackage[]{microtype}
  \UseMicrotypeSet[protrusion]{basicmath} 
}{}
\makeatletter
\@ifundefined{KOMAClassName}{
  \IfFileExists{parskip.sty}{%
    \usepackage{parskip}
  }{
    \setlength{\parindent}{0pt}
    \setlength{\parskip}{6pt plus 2pt minus 1pt}}
}{
  \KOMAoptions{parskip=half}}
\makeatother
\makeatletter
\ifx\paragraph\undefined\else
  \let\oldparagraph\paragraph
  \renewcommand{\paragraph}{
    \@ifstar
      \xxxParagraphStar
      \xxxParagraphNoStar
  }
  \newcommand{\xxxParagraphStar}[1]{\oldparagraph*{#1}\mbox{}}
  \newcommand{\xxxParagraphNoStar}[1]{\oldparagraph{#1}\mbox{}}
\fi
\ifx\subparagraph\undefined\else
  \let\oldsubparagraph\subparagraph
  \renewcommand{\subparagraph}{
    \@ifstar
      \xxxSubParagraphStar
      \xxxSubParagraphNoStar
  }
  \newcommand{\xxxSubParagraphStar}[1]{\oldsubparagraph*{#1}\mbox{}}
  \newcommand{\xxxSubParagraphNoStar}[1]{\oldsubparagraph{#1}\mbox{}}
\fi
\makeatother

\usepackage{longtable,booktabs,array}
\usepackage{calc} 
\usepackage{etoolbox}
\makeatletter
\patchcmd\longtable{\par}{\if@noskipsec\mbox{}\fi\par}{}{}
\makeatother
\IfFileExists{footnotehyper.sty}{\usepackage{footnotehyper}}{\usepackage{footnote}}
\makesavenoteenv{longtable}
\usepackage{graphicx}
\makeatletter
\newsavebox\pandoc@box
\newcommand*\pandocbounded[1]{
  \sbox\pandoc@box{#1}%
  \Gscale@div\@tempa{\textheight}{\dimexpr\ht\pandoc@box+\dp\pandoc@box\relax}%
  \Gscale@div\@tempb{\linewidth}{\wd\pandoc@box}%
  \ifdim\@tempb\p@<\@tempa\p@\let\@tempa\@tempb\fi
  \ifdim\@tempa\p@<\p@\scalebox{\@tempa}{\usebox\pandoc@box}%
  \else\usebox{\pandoc@box}%
  \fi%
}
\def\fps@figure{htbp}
\makeatother

\providecommand{\tightlist}{%
  \setlength{\itemsep}{0pt}\setlength{\parskip}{0pt}}

\usepackage[style=authoryear,backend=biber,maxcitenames=2,doi=true,isbn=false,eprint=false]{biblatex}
\ifPDFTeX
  \usepackage{upgreek}
  \newcommand\symbfupphi{\boldsymbol{\upphi}}
\else 
  \newcommand\symbfupphi{\symbfup{\phi}}
\fi

\DeclareMathOperator{\IR}{\mathbb{R}}

\let\geq\geqslant
\let\leq\leqslant
\let\epsilon\varepsilon

\usepackage{bm}

\usepackage{booktabs}
\usepackage{caption}
\usepackage{longtable}
\usepackage{colortbl}
\usepackage{array}
\usepackage{anyfontsize}
\usepackage{multirow}
\makeatletter
\@ifpackageloaded{caption}{}{\usepackage{caption}}
\AtBeginDocument{%
\ifdefined\contentsname
  \renewcommand*\contentsname{Table of contents}
\else
  \newcommand\contentsname{Table of contents}
\fi
\ifdefined\listfigurename
  \renewcommand*\listfigurename{List of Figures}
\else
  \newcommand\listfigurename{List of Figures}
\fi
\ifdefined\listtablename
  \renewcommand*\listtablename{List of Tables}
\else
  \newcommand\listtablename{List of Tables}
\fi
\ifdefined\figurename
  \renewcommand*\figurename{Figure}
\else
  \newcommand\figurename{Figure}
\fi
\ifdefined\tablename
  \renewcommand*\tablename{Table}
\else
  \newcommand\tablename{Table}
\fi
}
\@ifpackageloaded{float}{}{\usepackage{float}}
\floatstyle{ruled}
\@ifundefined{c@chapter}{\newfloat{codelisting}{h}{lop}}{\newfloat{codelisting}{h}{lop}[chapter]}
\floatname{codelisting}{Listing}

\makeatother
\makeatletter
\@ifpackageloaded{caption}{}{\usepackage{caption}}
\@ifpackageloaded{subcaption}{}{\usepackage{subcaption}}
\makeatother
\usepackage{bookmark}
\IfFileExists{xurl.sty}{\usepackage{xurl}}{} 
\hypersetup{
  pdftitle={How to optimize dynamic borrowing in basket trials -- A utility-based framework and results of a comparison study},
  pdfauthor={Lukas D Sauer; Alexander Ritz; Meinhard Kieser},
  pdfkeywords={basket trials, master protocols, utility
functions, optimization, Bayesian statistics, oncology, rare diseases},
  colorlinks=true,
  linkcolor={darkgray},
  filecolor={Maroon},
  citecolor={darkgray},
  urlcolor={darkgray},
  pdfcreator={LaTeX via pandoc}}

\title{How to optimize dynamic borrowing in basket trials -- A
utility-based framework and results of a comparison study}%
\usepackage[noblocks]{authblk}
\author[1]{Lukas D Sauer\thanks{\textsc{Correspondence.}\quad Lukas D
Sauer, Institute of Medical Biometry, Heidelberg University, Im
Neuenheimer Feld
130.3, 69120, Heidelberg, Germany. E-mail: \href{mailto:sauer@imbi.uni-heidelberg.de}{\nolinkurl{sauer@imbi.uni-heidelberg.de}}}}%
\author[2]{Alexander Ritz}%
\author[1]{Meinhard Kieser}%
\affil[1]{Institute of Medical Biometry, Heidelberg
University, Germany}%
\affil[2]{Institute of Mathematics, Clausthal University of
Technology, Germany}%
\date{August 1, 2026}%
\usepackage{etoolbox}
\makeatletter
\apptocmd{\maketitle}{\vspace*{-3em}}{}{}
\makeatother
\begin{document}
\maketitle
\begin{abstract}
\textbf{Background.} Basket trials are clinical trials in which one
treatment is investigated in multiple subpopulations within a single
trial. These subpopulations are referred to as strata in the following.
Such stratified patient cohorts are typical but not limited to early
phase oncological trials, where targeted cancer therapies can be
investigated in different strata defined by the tumor tissue. In terms
of statistical methodology, the stratification of the trial can be
leveraged using information borrowing methods. The idea is that the
strata will be analysed separately if the strata respond differently to
treatment, but that a stratum may share information from another if
their response rates are similar. This approach is used to increase
power while keeping type-I error inflation moderate.

\textbf{Methods.} A multitude of Bayesian and frequentist information
borrowing methods have been suggested in the past years. In order to
achieve good performance in terms of power and type-I error rate, the
borrowing methods can be tuned with respect to a set of possible outcome
scenarios. In this paper, we provide a framework for optimizing basket
trials, including a checklist of all relevant aspects that need to be
specified. The targeted compromise between power and type-I error rate
is defined by a utility function, which is then optimized using an
optimization algorithm. We investigated this framework by performing a
pre-specified comparison study.

\textbf{Results.} Part I of the comparison study compared optimization
algorithms in terms of reliability and efficiency, part II compared the
statistical performance of utility functions across several simulated
outcome scenarios. This study shows how targeting different utility
functions based on measures such as experiment-wise power, local power
in each stratum, or the expected number of correct decisions results in
different borrowing behavior of the optimized design. Furthermore, we
discuss a mathematical counterexample which shows that no uniformly most
powerful (UMP) test exists for basket trials, thus demonstrating the
limitations of optimized basket trials.

\textbf{Conclusions.} While information borrowing may result in power
gains for some scenarios, there is no design which is optimal for all
possible scenarios. Hence, a transparent optimization procedure is
crucial when planning basket trials.
\end{abstract}

\section{Background}\label{sec-introduction}

Master protocols are flexible clinical trial designs in which one or
multiple treatments can be investigated in one or multiple patient
populations within a single overarching trial structure. The term master
protocol encompasses a variety of different trial designs, among which
are the so called \emph{basket trials}: A basket trial is a single
clinical trial in which one treatment is investigated in several patient
populations. These designs are common in early phase oncological drug
trials \autocite{kasim_basket_2023}. In these settings, innovative drugs
such as monoclonal antibodies or immune checkpoint inhibitors are often
targeted at specific genetic tumor traits and can hence be applied in a
variety of tissues. For example, the INTUITT-NF2 trial was a phase 2,
multi-center basket trial which investigated the effect of brigatinib in
different progressive tumor types, namely vestibular schwannomas,
nonvestibular schwannomas, meningiomas, and ependymomas
\autocite{plotkin_brigatinib_2024}. The metaphoric name ``basket trial''
comes from the idea of collecting everything of interest in one basket,
i.e.~investigating all subpopulations in a single trial
\autocite{woodcock_master_2017}. The different subpopulations will be
referred to as \emph{strata} in the following; in the literature the
strata themselves are sometimes also called baskets. Basket trials are
often single-arm trials without a control group. Their endpoints are
often binary variables such as the objective radiographic response
(ORR). Basket trial designs are not only applied in oncology, but also
in other stratified patient populations with small sample sizes such as
in the context of rare diseases \autocite{khazen_basket_2025}.

From a biostatistical viewpoint, the stratified structure of basket
trials is both a challenge and an opportunity. If one analyses all
strata separately, the small sample sizes will result in rather low
power, i.e.~a low probability of detecting strata which are
\emph{active}, meaning they truly respond to treatment. If one pools the
data of all strata, the power will be increased, but one may have a
higher type-I error rate (TOER), i.e.~a higher probability of
erroneously rejecting the null hypothesis in strata which are
\emph{inactive}, meaning they truly do not respond to treatment. As a
compromise between these two options, a large variety of information
borrowing methods based on Bayesian and frequentist approaches has been
suggested for basket trial designs \autocite{pohl_categories_2021}.
These methods dynamically let strata borrow more information from other
strata if their response rates are similar, and analyse strata
separately if their response rates are heterogeneous. If one has
selected a statistical model for a basket trial design, one needs to
tune the performance of this model in order to achieve the optimal
compromise between power gains and inflation of TOER. In this paper, we
systematically investigate the process of optimizing a basket trial
design in a clearly structured framework. In order to investigate the
effect of our framework in optimizing basket trial designs, we conducted
a comparison study which was pre-specified in a study protocol. The
protocol was made public on arXiv in 2024 and later published in a
peer-reviewed journal \autocite{sauer_utilitybased_2025}.

The paper is structured as follows: In Section~\ref{sec-checklist}, we
provide a checklist of all aspects that need to be considered when
optimizing a basket trial design using our framework. In
Section~\ref{sec-basket-optim}, we discuss the mathematical foundations
of optimizing basket trial designs. In
Section~\ref{sec-utility-functions}, we define utility functions which
can be used as targets when optimizing basket trial designs, and in
Section~\ref{sec-algorithms}, we provide a list of different algorithms
which are candidates for optimizing basket trial designs.
Section~\ref{sec-protocol} provides a summary of the study protocol of
our comparison study. In Section~\ref{sec-results}, we present the
results of the comparison study. In Section~\ref{sec-discussion}, we
provide a mathematical counterexample that illustrates the limitations
of optimizing basket trial designs. With this example in mind, the
results of the comparison study are discussed and put into context. The
article ends with a brief conclusion in Section~\ref{sec-conclusion}.

\section{Methods}\label{methods}

\subsection{A checklist for optimizing basket
trials}\label{sec-checklist}

A basket trial design is a clinical trial design testing one treatment
in multiple strata. Planning basket trial designs is complex, hence the
different considerations in their design should be stated explicitly,
ideally following a clear framework. In
\textcite{kaizer_statistical_2021}, the authors discuss different
aspects of such a framework: In particular, they address the choice of
an analysis strategy, different statistical operating characteristics,
and the definition of a sensible set of outcome scenarios. They go forth
and describe different approaches to weight the importance of scenarios.
Building on their work, we suggest the following framework consisting of
aspects that should explicitly be addressed when planning the
statistical aspects of a basket trial design:

\begin{itemize}
\item
  \emph{Trial-specific aspects:}

  \begin{itemize}
  \tightlist
  \item
    \emph{Number of strata:} How many different subgroups/indications
    \(i = 1,\ldots,I\) are going to be investigated in the basket trial
    design?
  \item
    \emph{Endpoint type:} Is the investigated primary endpoint binary,
    categorical, ordinal, continuous or a time-to-event endpoint? Most
    published methods for basket trials consider binary endpoints,
    i.e.~response to treatment ``yes'' or ``no''. This article will
    chiefly discuss binary endpoints, too.
  \item
    \emph{Outcome scenario set:} Which treatment effects are to be
    expected in the different strata? Response scenarios differ
    depending on the respective endpoint type. In this article, we will
    mainly look at binary endpoints under a frequentist paradigm. In
    that setting, one needs to define response rates \(p_{0,i}\) for
    ``inactive'' strata \(i\), i.e.~such strata in which the treatment
    is deemed ineffective, and response rates \(p_{1,i}\) for ``active
    strata'', i.e.~such strata in which the treatment is deemed
    effective. Defining a scenario set then boils down to defining
    several response vectors
    \(\mathbf{p}=(p_{1}, \ldots,p_{i},\ldots, p_{I})\). One of these
    vectors should be the global null scenario in which all responses
    are the null response \(p_i=p_{0,i}\) and of course, there should be
    one or more alternative scenarios in which at least one stratum is
    active. When one has selected a set of scenarios
    \(\{\mathbf{p},\ldots\}\), one should also define how these
    scenarios are to be weighted when optimizing the basket trial by
    defining weights \(w_{\mathbf{p}}\) for each scenario set. In the
    following, we will only consider equal weights
    \(w_{\mathbf{p}}=\frac{1}{\#\{\mathbf{p},\ldots\}}\) with
    \(\sum_{\mathbf{p}\in\{\mathbf{p},\ldots\}}w_{\mathbf{p}}=1\). Other
    weights can be considered: \textcite{kaizer_statistical_2021}
    discuss weights depending on the number of active strata in the
    scenario and more complicated weighting schemes in which different
    weights are used in the calculation of average power and average
    type-I error. In a fully Bayesian setting, one would define prior
    distributions for each \(p_i\) instead of null and alternative
    scenarios with different weights.
  \end{itemize}
\item
  \emph{Operating characteristics:} Which operating characteristics are
  important in measuring the statistical performance of a basket trial
  design? Some operating characteristics are ``targets'', meaning that
  they are to be maximized when optimizing the trial design. Others are
  ``constraining'' meaning they should be minimized or at least below a
  certain threshold. Denote by \(H_0^i\) the null hypothesis in a
  stratum \(i\). Here are some typical operating characteristics:

  \begin{itemize}
  \tightlist
  \item
    \emph{Target operating characteristics:} marginal power in the
    single strata (i.e.~the probability to rightly reject the null
    hypothesis \(H_0^i\) in a truly active stratum \(i\)),
    experiment-wise power (EWP, i.e.~the probability to rightly reject
    \(H_0^i\) in at least one truly active stratum \(i\) in the trial),
    expected number of correct decisions (i.e.~the number of correct
    test decisions, that is rejections of \(H_0^i\) in truly active
    strata \(i\) and keeping \(H_0^i\) in truly inactive strata \(i\))
  \item
    \emph{Constraining operating characteristics:} sample size \(n_i\),
    marginal type-I error rate (TOER, i.e.~the probability to wrongly
    reject the null hypothesis \(H_0^i\) in a truly inactive stratum),
    family-wise error rate (FWER, i.e.~the probability to wrongly reject
    \(H_0^i\) in at least one truly inactive stratum in the trial)
  \end{itemize}

  \textcite{kaizer_statistical_2021} discuss different notions of
  defining FWER control in a basket trial: weak control means FWER is
  only to be controlled in the global null scenario, strong control in
  the sense of \textcite{dmitrienko_multiple_2009} means that FWER is to
  be controlled in every configuration of null and alternative
  scenarios, and strong control in the sense of
  \textcite{kaizer_statistical_2021} means FWER is to be controlled in a
  set of plausible scenarios. In a fully Bayesian setting, one may
  consider using assurance, i.e.~the Bayesian unconditional prior
  statistical power, instead of the frequentist power
  \autocite{ohagan_assurance_2005}.
\item
  \emph{Analysis strategy:} Should the different strata be evaluated in
  one pooled analysis, in completely separate analyses, or using a
  dynamic information borrowing approach? This is a difficult decision.
  While methodological research is focused on novel information
  borrowing methods, these methods have rarely been applied in actual
  trial planning \autocite{hobbs_basket_2022}. A recent scoping review
  found 12 trials which used dynamic borrowing between their arms
  \autocite{weru_uptake_2026}. A full discussion of the reasons for the
  slow adoption of information borrowing methods is beyond the scope of
  this paper, but one methodological challenge is discussed in
  Section~\ref{sec-counterex}.
\item
  \emph{Specification of information borrowing component (if borrowing
  is desired):} If one has come to the conclusion that an information
  borrowing component is justified for the basket trial at hand, one
  needs to further specify the components of this borrowing component.
  This includes several aspects:

  \begin{itemize}
  \tightlist
  \item
    \emph{Statistical design:} One needs to specify a statistical
    mechanism which implements the information borrowing. There are
    frequentist mechanisms (e.g.~clustering and pooling) and Bayesian
    mechanisms (e.g.~based on the Bayesian hierarchical model or on the
    beta-binomial model). A detailed classification of Bayesian
    mechanisms can be found in \textcite{pohl_categories_2021}.
    Simulation studies in order to compare different Bayesian mechanisms
    have been conducted
    \autocite{broglio_comparison_2022,baumann_basket_2024a}. Most
    borrowing mechanisms have tuning parameters which need to be adapted
    to the trial's characteristics. The optimal amount of borrowing
    depends on sample size, number of strata, response scenarios of
    interest, and response rates, as we will demonstrate in several
    examples below.
  \item
    \emph{Optimization approach for tuning the design:} With respect to
    the scenario set defined above, one needs to define an optimization
    approach which will tune the statistical design parameters in order
    to maximize target operating characteristics while respecting
    constraints on constraining operating characteristics.

    \begin{itemize}
    \tightlist
    \item
      \emph{Analytical optimization:} If the calculation of the
      performance measures has a nice analytical form, one may apply
      analytical optimization techniques such as Lagrange multipliers
      or, even simpler, finding roots of the derivative.
    \item
      \emph{Numerical optimization:} If analytical optimization is not
      possible, define a numerical approach with the following
      components:

      \begin{itemize}
      \tightlist
      \item
        \emph{Utility function:} A function which takes the design's
        tuning parameters as inputs and outputs some expression of the
        target and/or constraining performance measures.
      \item
        \emph{Optimization algorithm:} An algorithm to optimize the
        utility function.
      \end{itemize}
    \end{itemize}
  \end{itemize}
\item
  \emph{Adaptive component:} Should the trial encompass interim analyses
  for efficacy or futility, and, potentially, the possibility for sample
  size recalculation? What are the decision rules for the stops and for
  the sample size recalculation? The former two are discussed in
  \textcite{pohl_categories_2021}. So far, we do not know of a
  discussion of sample size recalculation in the context of basket
  trials.
\end{itemize}

The components of the framework described above can be used as a
checklist when planning a basket trial. For example, we could fill the
checklist as follows:

\begin{itemize}
\tightlist
\item
  \emph{Trial-specific aspects:}

  \begin{itemize}
  \tightlist
  \item
    \emph{Number of strata:} \(I=3\)
  \item
    \emph{Endpoint type:} binary (response yes/no)
  \item
    \emph{Scenario set:} \((0.2, 0.2, 0.2)\), \((0.2, 0.2, 0.5)\),
    \((0.2, 0.5, 0.5)\), and \((0.5, 0.5, 0.5)\), where \(p_0 = 0.2\) is
    the null response rate
  \end{itemize}
\item
  \emph{Operating characteristics:}

  \begin{itemize}
  \tightlist
  \item
    \emph{Target operating characteristics:} expected number of correct
    decisions (ECD)
  \item
    \emph{Constraining operating characteristics:} family-wise error
    rate (FWER)
  \end{itemize}
\item
  \emph{Analysis strategy:} Analysis with information borrowing.
\item
  \emph{Specification of information borrowing component:}

  \begin{itemize}
  \tightlist
  \item
    \emph{Statistical design:} Beta-binomial borrowing design from
    \textcite{fujikawa_bayesian_2020}
  \item
    \emph{Optimization approach for tuning the design:}

    \begin{itemize}
    \tightlist
    \item
      \emph{Utility function:} \texttt{u\_ecd\_avg\_pen} (defined in
      Section~\ref{sec-utility-functions} below, resulting in moderate
      borrowing)
    \item
      \emph{Optimization algorithm:} grid search
    \end{itemize}
  \end{itemize}
\item
  \emph{Adaptive component:} No interim analyses.
\end{itemize}

As far as we know, there has not been a general discussion of the
optimization approach of basket trial designs. A detailed investigation
into the optimization of basket trial designs is the topic of this
paper. In particular, the rest of this paper will address two questions:

\begin{itemize}
\tightlist
\item
  Part I: Which optimization algorithm as defined in
  Section~\ref{sec-algorithms} is efficient and reliable in finding the
  optimal tuning parameter combination?
\item
  Part II: Which utility function as defined in
  Section~\ref{sec-utility-functions} is appropriate for tuning the
  parameters and how do they compare to the untuned default parameters?
\end{itemize}

Results of part I and II can be found in
Section~\ref{sec-perf-algorithms} and
Section~\ref{sec-perf-utility-functions}, respectively. In order to
address these questions, we planned a comparison study, which was
pre-specified in a study protocol published in
\textcite{sauer_utilitybased_2025}. In the study protocol, we
differentiated between part II ``Which utility function is
appropriate?'' and part III ``How do the utility functions' optimal
results compare to the default parameters?''. As these questions are
answered with the same performance measures in the same figures, we will
henceforth only refer to part II for both of these questions.

\subsection{Basket trial designs as constrained optimization
problems}\label{sec-basket-optim}

From a statistical perspective, a basket trial design is a decision
function which gives a test decision for each stratum based on the data
for all strata, i.e.~a function \(D(\cdot)\) defined as follows:

\[ D: \mathcal X \to \{0,1\}^I, \quad\text{data}\mapsto D(\text{data})_i=\begin{cases} 1 & \text{test rejects $H_0^i$ in stratum }i, \\ 0 & \text{else.} \end{cases} \]

Usually a basket trial design is subject to a vector of tuning
parameters \(\symbfupphi\in\IR\), i.e.
\(D(\cdot)=D_{\symbfupphi}(\cdot)\). The task of optimizing a basket
trial design means finding a solution of the following constrained
optimization problem,

\begin{align*}
\text{find \quad} & \symbfupphi^* =\arg\max_{\symbfupphi} u(D_{\symbfupphi})  \\
\text{subject to \quad} & v(D_{\symbfupphi}) \leq c\\
\end{align*}

where \(u(\cdot)\) is a utility function to be maximized, and
\(v(\cdot)\) is a constraint function. For optimization with a numeric
optimization algorithm, it can be useful to find an alternative
formulation which incorporates the constraints into the definition of
\(u(\cdot)\), e.g.~by imposing a penalty whenever the constraints are
violated.

\subsection{Utility functions}\label{sec-utility-functions}

When optimizing analysis methods for clinical trials, the most important
measure of efficiency is usually power, i.e.~the probability to declare
a truly effective treatment as effective. The constraints are usually
based on type-I error, i.e.~the probability to wrongly declare an
ineffective treatment as effective. As explained above, we aim to
optimize these aspects using a utility function. In the literature on
basket trials, one finds different definitions of utility functions, two
of which are considered in this comparison study:
\textcite{broglio_comparison_2022} used the ECD,
\textcite{jiang_optimal_2021} summed over the power in each active
stratum and subtracted the type-I error in each inactive stratum and
averaged this quantity across a set of different scenarios. Their
approaches correspond to the functions \texttt{u\_ecd\_avg} and
\texttt{u\_2pow\_avg} as explained in the following.

In our comparison study, we extended these approaches by looking at
twelve different utility functions as shown in
Table~\ref{tbl-utility-functions}. The utility functions \texttt{u\_ewp}
and \texttt{u\_ecd} target EWP and ECD, respectively, while returning
the negative FWER under the global null hypothesis if it exceeds a
pre-defined threshold. The utility functions \texttt{u\_2ewp} and
\texttt{u\_2pow} target the difference of EWP minus FWER and local power
minus local TOER, respectively, with an additional penalty if the type-I
error exceeds a predefined threshold. Each of these utility functions
can be averaged across the set of all relevant scenarios, resulting in
four more utility functions denoted with the suffix \texttt{\_avg}.
Finally, in addition to averaging across a set of scenarios, one can
also check whether the local TOER in any stratum in any scenario exceeds
a threshold which is deemed unacceptable, \(\eta_3=0.2\) in the example
at hand. If this is the case, the maximal local TOER multiplied by -1000
is returned, causing the algorithm to search for parameters which
respect this threshold. This results in another four utility functions
denoted with the suffix \texttt{\_avg\_pen}.

\begin{table}

\caption{\label{tbl-utility-functions}Utility functions}

\centering{

\begin{tabular*}{\linewidth}{@{\extracolsep{\fill}}>{\raggedright\arraybackslash}p{\dimexpr 90.00pt -2\tabcolsep-1.5\arrayrulewidth}|>{\raggedright\arraybackslash}p{\dimexpr 341.25pt -2\tabcolsep-1.5\arrayrulewidth}}
\toprule
\textbf{Abbreviation} & \textbf{Definition} \\ 
\midrule\addlinespace[2.5pt]
\texttt{u\_ewp} & \textbf{Discontinuous family-wise power-error function}\medbreak $u_{\text{ewp}}(\symbfupphi,\mathbf{p},\mathbf{p_0})=\begin{cases}\mathrm{ewp}(\mathbf{p}) & \text{if } \mathrm{fwer}(\mathbf{p_0}) < \eta_1,\\ -\xi_1\cdot\mathrm{fwer}(\mathbf{p_0}) & \text{if } \mathrm{fwer}(\mathbf{p_0}) \geq \eta_1,\end{cases}$ \medbreak with threshold $\eta_1=0.05$ and penalty $\xi_1=1$\smallbreak ~ \\ 
\texttt{u\_ecd} & \textbf{Expected number of correct decisions} \medbreak $u_{\text{ecd}}(\symbfupphi,\mathbf{p},\mathbf{p_0}) = \begin{cases}\mathrm{ecd}(\mathbf{p}), & \text{if } \mathrm{fwer}(\mathbf{p_0}) < \eta_1,\\ -\xi_1\cdot\mathrm{fwer}(\mathbf{p_0}) & \text{if } \mathrm{fwer}(\mathbf{p_0}) \geq\eta_1,\end{cases}$ \medbreak with threshold $\eta_1=0.05$ and penalty $\xi_1=1$\smallbreak ~ \\ 
\texttt{u\_2ewp} & \textbf{Two-level family-wise power-error function}\medbreak $u_{\text{2ewp}}(\symbfupphi,\mathbf{p}) = \mathrm{ewp}(\mathbf{p}) - \left(\xi_1\mathrm{fwer}(\mathbf{p})+\xi_2(\mathrm{fwer}(\mathbf{p}) - \eta_2)\mathbf 1(\mathrm{fwer}(\mathbf{p}) - \eta_2) \right)$ \medbreak with threshold $\eta_2=0.1$ and penalty $\xi_1=1$\smallbreak ~ \\ 
\texttt{u\_2pow} & \textbf{Two-level stratum-wise power-error function}\medbreak $u_{\text{2pow}}(\symbfupphi,\mathbf{p}) = \sum_{i\in R}\mathrm{pow}_i(\mathbf{p}) - \sum_{j\in R^c}(\xi_1\mathrm{toer}_j(\mathbf{p}) + \xi_2(\mathrm{toer}_j(\mathbf{p}) - \eta_2) \mathbf 1(\mathrm{toer}_j(\mathbf{p}) - \eta_2) ),$ \medbreak with threshold $\eta_2=0.1$ and penalty $\xi_1=1$\smallbreak ~ \\ 
\texttt{u\_ewp\_avg}, \texttt{u\_ecd\_avg}, \texttt{u\_2ewp\_avg}, \texttt{u\_2pow\_avg} & \textbf{Scenario-averaged utility functions}\medbreak $\bar{u_l}(\symbfupphi,\mathbf{p_0})=\sum_{\mathbf p\in\{\mathbf{p},\ldots\}}w_\mathbf{p}u_l( \mathbf{p}, \mathbf{p_0})$ \medbreak with weights $w_\mathbf{p}=\frac{1}{\#\{\mathbf{p},\ldots\}}$\smallbreak ~ \\ 
\texttt{u\_ewp\_avg\_pen}, \texttt{u\_ecd\_avg\_pen}, \texttt{u\_2ewp\_avg\_pen}, \texttt{u\_2pow\_avg\_pen} & \textbf{Scenario-averaged utility functions with penalty of maximal TOER inflation}\medbreak $\bar{u}_{l,\text{pen}}(\symbfupphi,\mathbf{p_0})=\begin{cases}\bar u_l( \mathbf{p_0})& \text{if }\max_{\mathbf p, j} \mathrm{toer}_j(\mathbf{p}) < \eta_3,\\-\xi_3\cdot\max_{\mathbf p, j} \mathrm{toer}_j(\mathbf{p})& \text{if }\max_{\mathbf p, j} \mathrm{toer}_j(\mathbf{p}) \geq \eta_3,\end{cases}$ \medbreak with threshold $\eta_3=0.2$ and penalty $\xi_3=1000$ \\ 
\bottomrule
\end{tabular*}
\begin{minipage}{\linewidth}
\vspace{.05em}
\footnotesize\textsc{Symbols.} $R$ and $R^c\subseteq \{1,\ldots,I\}$ are the sets of truly active and inactive strata, respectively. The vector $\mathbf{p}=(p_1,\ldots,p_I)$ is a response scenario, i.e. a vector of true response rates for each stratum. $\mathbf{p_0}$ is the response scenario under the global null hypothesis. $\mathrm{pow}_i(\cdot)$ and $\mathrm{toer}_i(\cdot)$ are the local power and local type-I error probability, i.e. the probability of rejecting the null hypothesis in stratum $i$ given that the stratum is active or inactive, respectively. $\mathrm{ewp}(\cdot)$ and $\mathrm{fwer}(\cdot)$ are the experiment-wise power and the family-wise error rate, i.e. the probability of rejecting at least one null hypothesis in a truly active stratum/in a truly inactive stratum, respectively. $\symbfupphi$ is a vector of tuning parameters of the basket trial design.\\
\end{minipage}

}

\end{table}%

\subsection{Optimization algorithms}\label{sec-algorithms}

For the optimization of a basket trial design's tuning parameters, one
could theoretically use any optimization algorithm that allows for some
constraints on the parameters. We considered a set of different
optimization algorithms. A detailed description of their implementation
can be found in the simulation protocol \autocite[section
7]{sauer_utilitybased_2025}. Here, we provide a brief summary.

\emph{Grid search} is the simplest in terms of implementation,
evaluating the utility function on a pre-defined grid of parameter
combinations. Then, \emph{simulated annealing} is a meta-heuristic
optimization algorithm based on the Metropolis-Hastings algorithm. A
so-called temperature schedule ``cools down'' the parameter space during
the course of the procedure. Temperature is a parameter that defines the
probability of the algorithm proceeding to a state that is worse than
the current one, which will in turn allow the algorithm to escape local
minima. We looked at the classic algorithm without boundaries on the
parameter space suggested by \autocite{kirkpatrick_optimization_1983},
where we manually implemented boundaries by setting the utility function
to \(-\infty\). Furthermore, we looked at a variant of the algorithm,
where the algorithm itself controls boundaries on the parameter space by
reflecting values which cross the boundaries
\autocite{haario_simulated_1991}. Different start temperatures \(T\) of
the algorithm were tested. Furthermore, we looked at \emph{constrained
optimization by linear approximation} (COBYLA), which linearly
approximates the shape of the utility function by simplices in order to
generate the next candidate of the search \autocite{powell_direct_1994}.
Finally, we looked at two nature-inspired meta-heuristic algorithms:
\emph{differential evolution} (DE) is based on the idea of combining
several candidates to new candidates similar to how genetic evolution
and mutation works \autocite{das_differential_2011}. Finally, the
\emph{particle swarm optimizer} (PSO) mimics the behavior of a swarm of
animals. The variant of the PSO that we looked at is called ``grey wolf
optimizer''. However, it has meanwhile come to our attention that the
grey wolf optimizer is actually a variant of PSO with little innovative
value \autocite{camacho-villalon_exposing_2023}, hence we use the term
PSO in the following.

The grid search algorithm and the constrained simulated annealing
algorithm were implemented in the R package \emph{optimizr} written for
this comparison study \autocite{sauer_optimizr_2026}. An implementation
of unconstrained simulated annealing can be found in the \emph{stats}
package which is part of the R core \autocite{r_2026}, COBYLA can be
found in the \emph{nloptr} package \autocite{johnson_nlopt_2008}, and DE
and PSO ``grey wolf'' can be found in the \emph{metaheuristicOpt}
package \autocite{septemriza_metaheuristic_2019}.

\section{Comparison study}\label{sec-protocol}

\subsection{Outcome scenario sets}\label{sec-scenario-sets}

The performance of the utility functions in finding the optimal
parameter vector was investigated in seven different sets of scenarios,
each set representing possible outcomes of a basket trial with a fixed
number of strata and a fixed number of patients per stratum. Some of the
scenario sets were inspired by other methodological papers on basket
trials
\autocite{baumann_basket_2024a,fujikawa_bayesian_2020,krajewska_new_2021},
others were inspired by actual basket trials reported in a systematic
review on basket trials
\autocite{kasim_basket_2023,gileadsciences_study_2024}.

The scenario sets are defined in Table~\ref{tbl-scenario-sets}. The
majority of scenario sets considers simple scenarios consisting of a
null response rate \(p_0\), an alternative response rate, and a number
of active strata \(a=0,\ldots, I\), i.e.~the number of those strata
whose true response rate is the alternative response rate. Only for the
scenario with \(I=4\) strata denoted by \texttt{sc\_bau04}, we also
considered two more complex response scenarios, a linear increase across
the different strata (``linear'', number of active strata \(a=3\)), and
a global alternative scenario in which all strata are active but have
different response rates (``one in the middle'', number of active strata
\(a=4\)).

\begin{table}

\caption{\label{tbl-scenario-sets}Outcome scenario sets}

\centering{

\begin{tabular*}{\linewidth}{@{\extracolsep{\fill}}>{\raggedright\arraybackslash}p{\dimexpr 56.25pt -2\tabcolsep-1.5\arrayrulewidth}|>{\raggedleft\arraybackslash}p{\dimexpr 18.75pt -2\tabcolsep-1.5\arrayrulewidth}>{\raggedleft\arraybackslash}p{\dimexpr 18.75pt -2\tabcolsep-1.5\arrayrulewidth}>{\raggedleft\arraybackslash}p{\dimexpr 26.25pt -2\tabcolsep-1.5\arrayrulewidth}>{\raggedleft\arraybackslash}p{\dimexpr 30.00pt -2\tabcolsep-1.5\arrayrulewidth}>{\raggedright\arraybackslash}p{\dimexpr 150.00pt -2\tabcolsep-1.5\arrayrulewidth}>{\raggedright\arraybackslash}p{\dimexpr 112.50pt -2\tabcolsep-1.5\arrayrulewidth}}
\toprule
\textbf{Abbreviation} & \(I\) & \(n_i\) & \(n_{\text{total}}\) & \(p_0\) & \textbf{Scenarios} & \textbf{Reference} \\ 
\midrule\addlinespace[2.5pt]
\texttt{sc\_fuj03} & 3 & 24 & 72 & 0.20 & $\mathbf{p}=(0.20,\ldots, 0.20, \underbrace{0.50,\ldots, 0.50}_a)$ with $0\leq a \leq 3$ & \cite{fujikawa_bayesian_2020} \\ 
\texttt{sc\_bau04} & 4 & 20 & 80 & 0.15 & $\mathbf{p}=(0.15,\ldots, 0.15, \underbrace{0.40,\ldots, 0.40}_a)$ with $0\leq a \leq 4,$
                
$\mathbf{p}=(0.4, 0.4, 0.3, 0.5)$ 
                
"one in the middle",

$\mathbf{p}=(0.15, 0.25, 0.35, 0.45)$ 

"linear" & \cite{baumann_basket_2024a} \\ 
\texttt{sc\_med04} & 4 & 36 & 144 & 0.10 & $\mathbf{p}=(0.10,\ldots, 0.10, \underbrace{0.35,\ldots, 0.35}_a)$ with $0\leq a \leq 4$ & \cite{gileadsciences_study_2024} (NCT01848834) \\ 
\texttt{sc\_lrg03} & 3 & 54 & 162 & 0.15 & $\mathbf{p}=(0.15,\ldots, 0.15, \underbrace{0.30,\ldots, 0.30}_a)$ with $0\leq a \leq 3$ & \cite{kasim_basket_2023} (NCT01631552) \\ 
\texttt{sc\_kra08} & 8 & 15 & 120 & 0.15 & $\mathbf{p}=(0.15,\ldots, 0.15, \underbrace{0.45,\ldots, 0.45}_a)$ with $0\leq a\leq 8$ & \cite{krajewska_new_2021} \\ 
\texttt{sc\_med09} & 9 & 23 & 207 & 0.01 & $\mathbf{p}=(0.01,\ldots, 0.01, \underbrace{0.10,\ldots, 0.10}_a)$ with $0\leq a\leq 9$  & \cite{kasim_basket_2023} (NCT02454972) \\ 
\texttt{sc\_lrg20} & 20 & 24 & 480 & 0.10 & $\mathbf{p}=(0.10,\ldots, 0.10, \underbrace{0.35,\ldots, 0.35}_a)$ with $a = 0, 2, 4, \ldots, 20$ & \cite{kasim_basket_2023} (NCT02054806) \\ 
\bottomrule
\end{tabular*}
\begin{minipage}{\linewidth}
\vspace{.05em}
\footnotesize\textsc{Symbols.} $I$ number of strata, $n_i$ sample size per stratum, $n_{\text{total}}$ total sample size, $p_0$ response rate under null hypothesis, $\mathbf{p}$ outcome scenario vector, $a$ number of active strata.\\
\end{minipage}

}

\end{table}%

\subsection{Basket trial design under
consideration}\label{sec-basket-design}

The utility functions presented can be applied for the optimization of a
basket trial design that has tuning parameters for determining its
borrowing behavior. For investigation in our comparison study, we chose
the basket trial design developed by \textcite{fujikawa_bayesian_2020}.
The design is based on the Bayesian beta-binomial model. Let \(p_i\) be
the true proportion of responders in a stratum \(i\in {1,\ldots, I}\).
Suppose we observed \(r_i\) responders among the \(n_i\) recruited
patients. The usual posterior distribution of the beta-binomial model is
\[ p_i \sim \mathrm{Beta}(a_i + r_i, b_i + n_i - r_i),\]

where \(a_i, b_i>0\) are parameters of the prior distribution
\(p_i\sim\mathrm{Beta}(a_i, b_i)\). In the following, we will set the
prior parameters \(a_i=b_i=1\), which means that the prior distribution
is the uniform distribution on the interval \([0,1]\). In the design by
Fujikawa et al., the usual posterior is modified in order to borrow data
from other strata as well,

\begin{equation}\protect\phantomsection\label{eq-borrowing-post}{
p_i \sim \mathrm{Beta}\left(\sum_j\omega_{ij}\cdot(a_j + r_j), \sum_j\omega_{ij}\cdot(b_j + n_j - r_j)\right).
}\end{equation}

The weights are defined as \(\omega_{ii}=1\) and
\(\omega_{ij}=\mathbf{1}(\tilde\omega_{ij}^\varepsilon>\tau)\cdot\tilde\omega_{ij}^\varepsilon\)
for \(i\neq j\) with \(\tau\in[0,1]\) and \(\varepsilon>0\). The
similarity measure \(\tilde\omega_{ij}\) is defined as one minus the
Jensen-Shannon divergence of the usual posterior distribution in the
strata \(i\) and \(j\).

Calculating the Jensen-Shannon divergence (JSD) involves numerical
integration. As it only needs to be calculated once during the
optimization of the tuning parameters, this is usually not a
computational issue. If one requires a faster alternative, one can
replace the JSD by the Hellinger distance as we suggested in the study
protocol. This approach delivers very similar performance to the JSD
after optimization, see Figure~\ref{fig-appendix-global-hld-vs-jsd}.

To decide whether a stratum is active, the posterior probability of
being above a target response rate \(p_i^*\) must be above a threshold
\(\lambda\in[0,1]\), i.e. \[P(p_i>p_i^*|\text{data})\geq\lambda,\] where
the posterior probability is calculated using the posterior with
borrowing from Equation~\ref{eq-borrowing-post}.

In summary, Fujikawa's basket trial design has three tuning parameters:
\(\varepsilon\) and \(\tau\) determine the shape of the borrowing
weights, and \(\lambda\) is the decision threshold. Fujikawa et
al.~suggested two default parameter combinations, which we will refer to
as \texttt{fuj1} and \texttt{fuj2} in the following, namely
\((\lambda, \varepsilon, \tau)=(0.99, 2, 0)\) and \((0.99, 2, 0.5)\).
The effect of the tuning parameters on type-I error can be seen in
Figure~\ref{fig-appendix-reject-params}. The design is in use in an
oncological basket trial \autocite{kawamoto_phase_2024}; however, the
trial's study protocol does not provide details.

We investigated how optimizing this parameter vector
\(\symbfupphi=(\lambda, \varepsilon, \tau)\) affects the basket trial
design's performance characteristics in the comparison study that is
described in the following.

\subsection{Comparison protocol}\label{comparison-protocol}

In order to investigate the choice of an efficient and reliable
optimization algorithm as well as the choice of an appropriate utility
function, we set up a comparison study consisting of two parts. Some
calculations were performed using Monte Carlo simulation methods, others
could be calculated exactly up to the precision of numerical
integration.

Both parts were planned using the ADEMP scheme of simulation studies
\autocite{morris_using_2019}. Here, we provide a brief summary of the
comparison protocols for both parts. A detailed description can be found
in the comparison protocol \autocite{sauer_utilitybased_2025}.

\subsubsection{Part I: Performance of optimization
algorithms}\label{part-i-performance-of-optimization-algorithms}

Part I addressed the question of the choice on an efficient and reliable
optimization algorithm and consisted of the following aspects:

\begin{itemize}
\tightlist
\item
  Aim: Selection of the fastest reliable optimization algorithm for
  optimization of the parameters of Fujikawa's basket trial design.
\item
  Test problems: Two test problems were posed to each optimization,
  consisting of the optimization of the utility functions
  \texttt{u\_ecd\_avg} and \texttt{u\_2pow\_avg} in the scenario
  \texttt{sc\_bau04}. The deterministic algorithms (grid search, COBYLA)
  were executed once on each test problem; the stochastic algorithms
  were executed \(n_{\text{runs}}=50\) times.
\item
  Estimand/target: Finding the optimal parameter combination with
  respect to the utility function as fast as possible.
\item
  Methods: Eight different optimization algorithms as described in
  Section~\ref{sec-algorithms} above, namely grid search, COBYLA,
  bounded simulated annealing with start temperatures \(T=100\),
  \(T=10\), and \(T=1\), unbounded simulated annealing with start
  temperature \(T=10\), DE, and PSO.
\item
  Performance measures: Efficiency (number of utility function
  evaluations), user CPU time, system CPU time, wall clock time, memory
  usage), internal consistency (mean and minimal/maximal optimal utility
  function value, mean and minimal/maximal optimal parameters, marginal
  sample standard deviations of parameters, 95\%-confidence intervals of
  the means), external consistency (success of finding an optimal
  utility value greater than or equal to the grid search results,
  minimal and maximal difference and 95\%-confidence interval of the
  difference to the grid search result).
\item
  Reporting: Tabular reports of the performance measures including Monte
  Carlo standard errors (MCSEs) in case of simulated results, line plots
  of the convergence of simulated annealing, heat maps of the utility
  function values calculated by the grid search algorithm.
\end{itemize}

\subsubsection{Part II: Performance of utility
functions}\label{part-ii-performance-of-utility-functions}

Part II addressed the choice of an appropriate utility function and
consisted of the following aspects:

\begin{itemize}
\tightlist
\item
  Aim: Selection of the utility function which achieves the best
  compromise between single-stratum power and EWP on the one hand and
  single-stratum TOER and FWER on the other hand.
\item
  Data: Seven scenario sets with \(I=3\) to \(20\) strata as described
  in Section~\ref{sec-scenario-sets} in which the utility functions were
  optimized using the best algorithm from part I. The performance
  measures were either calculated exactly or using Monte Carlo methods
  as specified in Section~\ref{sec-software}.
\item
  Estimand/target: The optimal parameter vector with respect to that
  utility function and with respect to the performance measures below.
\item
  Methods: Twelve utility functions as described in
  Section~\ref{sec-utility-functions} and the default parameter vectors
  \texttt{fuj1} and \texttt{fuj2}.
\item
  Performance measures: marginal rejection rates in each stratum
  (i.e.~local TOER or local power), FWER, EWP, ECD, utility function
  values for all utility functions.
\item
  Reporting: Tabular reports and dot plots of all performance measures.
  Contrary to the protocol, we used line plots for the global
  performance measures as they show tendencies more clearly.
\end{itemize}

For comparison with a conservative frequentist evaluation method, we
also calculated the performance measures of part II for separate
binomial tests in each stratum, unadjusted for multiplicity. Their
results are denoted by \texttt{freq}.

\subsection{Software}\label{sec-software}

The full run of the comparison study was executed on the
high-performance computing cluster \emph{BinAC2} of the University of
Tübingen, Germany. This cluster is part of \emph{bwHPC}, the high
performance computing services of the German State of Baden-Württemberg.
The jobs were assigned up to 50 physical cores with a base frequency of
2.80 or 2.95 GHz and maximal RAM usage of 400 GB
\autocite{_binac2_2025}. All computations were programmed in R version
4.6.0 \autocite{r_2026}.

Three R packages were developed in the process of this comparison study:
\emph{optimizr} \autocite{sauer_optimizr_2026} provides implementations
of the optimization algorithms grid search and bounded simulated
annealing, \emph{baskwrap} \autocite{sauer_baskwrap_2026} provides a
unified interface to the R packages \emph{baskexact} and \emph{basksim}
\autocite{baumann_baskexact_2024a,baumann_basksim_2026}, which were used
for the calculation of the performance measures of Fujikawa's basket
trial design, and \emph{baskoptr} provides implementations of the
utility functions \autocite{sauer_baskoptr_2026}. All packages are
available on the Comprehensive R Archive Network (CRAN).

For two of the smaller scenario sets in terms of number of strata and
sample size, \texttt{sc\_fuj03} and \texttt{sc\_bau04}, the operating
characteristics of Fujikawa's design were calculated exactly using
numerical integration, for all others, they were approximated using
random draws from the binomial distribution defined by the respective
scenario sets. The number of draws was \(n_{\text{sim}}=1000\) during
the progress of the optimization algorithms. Once the optimal result was
found, the final performance measures were calculated once more with
\(n_{\text{sim}}=10000\) draws. Approximation of rejection rates using
Monte Carlo approximations is commonly applied in basket trials
\autocites[see
e.g.][]{broglio_comparison_2022,baumann_basket_2024a,asano_practical_2023a}.

\section{Results}\label{sec-results}

\subsection{Part I: Performance of optimization
algorithms}\label{sec-perf-algorithms}

In terms of computational efficiency measured by runtime, the grid
search algorithm was the fastest, taking 25 minutes for the optimization
of the \texttt{u\_ecd\_avg} function, followed by the COBYLA algorithm
with 113 minutes, and then the meta-heuristic algorithms simulated
annealing and PSO with runtimes ranging from 578 to 1,504 minutes, see
Figure~\ref{fig-appendix-alg-efficiency-u-ecd-avg}. The DE algorithm was
not tested as its implementation generated candidate parameter vectors
which were out of bounds, resulting in occasional error messages during
implementation. When optimizing \texttt{u\_2pow\_avg}, the results were
slightly less drastic, but still visibly showed the same trend, see
Figure~\ref{fig-appendix-alg-efficiency-u-2pow-avg}. One should remark
that the runtime benefit of the grid search is due to a parallelized
implementation on 50 cores. If fewer cores had been available, the
difference would have been less pronounced. While parallelization is
impossible for the sequential implementation of the meta-heuristic
algorithms, one could also adapt them to stop early as soon as a point
is reached where changes in the parameter space only result in tiny
improvements in the utility function. In fact, the COBYLA algorithm
employs such an early stopping rule which is the reason for its
relatively fast performance despite its sequential nature.

In terms of external validity, most algorithms did not reliably succeed
in finding parameter values with higher utility than the ones found by
the deterministic grid search algorithm. Only the bounded simulated
annealing with start temperature \(T=1\) succeeded in finding better
values of the \texttt{u\_ecd\_avg} than grid search in 38 of 50 runs
(76\%), see Figure~\ref{fig-appendix-alg-reliability-u-ecd-avg}.
However, this advantage did not persist in optimizing the
\texttt{u\_2pow\_avg} function (better values in 24\% of the runs, see
Figure~\ref{fig-appendix-alg-reliability-u-2pow-avg}). One reason for
this is the fact that large parts of the parameter space result in very
similar function values (i.e.~the utility function is ``flat'') which
makes it hard for these algorithms to find the optimum.

To summarize, grid search appeared to be the most useful algorithm in
terms of both runtime and its ability to find the optimal result.

In the protocol of the comparison study, we defined a decision rule for
the selection of the best optimization algorithm. Optimization
algorithms were to be subsequently compared by internal reliability,
external reliability and speed. The best optimization algorithm was to
be used in the next section for optimizing the utility functions in the
different scenarios. This decision rule was imprecisely defined as no
clear performance measures were defined for comparing internal and
external reliability. This vagueness turned out to be uncritical as the
results so clearly favored grid search. As no algorithm got close to
grid search in consistently finding the optimal results (``external
reliability''), we decided to use grid search for optimization in the
following section.

\subsection{Part II: Performance of utility
functions}\label{sec-perf-utility-functions}

In the following two sections, we will describe how the utility
functions performed in optimizing the basket trial with respect to
different performance measures.

We will first address the results in the example scenario set suggested
by Fujikawa and colleagues (\texttt{sc\_fuj03}) in detail. Afterwards,
we will look at how the performance in the other six scenario sets is
different or similar to \texttt{sc\_fuj03}.

\subsubsection{Performance of utility functions in Fujikawa's example
scenario}\label{sec-perf-sc-fuj03}

The scenario set \texttt{sc\_fuj03} is a basket trial with \(I=\) 3
strata with \(n =\) 24 patients per stratum. The null hypothesis is a
response rate of \(p_0 =\) 0.2. The four scenarios considered range from
the global null, i.e. \(\mathbf p = (0.2, 0.2, 0.2)\) to a global
alternative, i.e.~ \(\mathbf p = (0.5, 0.5, 0.5)\).

In order to allow for an overall comparison of the utility functions'
performance, we defined the following data-driven (!) classifications
based on the comparison study's results:

\begin{itemize}
\tightlist
\item
  \emph{Too conservative} (-C): The ECD in all scenarios except for the
  global null scenario is below the ECD of the separate frequentist
  tests.
\item
  \emph{Similar to frequentist design (+F)}: The ECD in all scenarios is
  equal to the ECD of the separate frequentist tests.
\item
  \emph{Conservative} (+C): The FWER in all scenarios is less or equal
  to \(1.5\cdot \mathrm{FWER}_0\), where
  \(\mathrm{FWER}_0=1 - (1 - 0.025)^I\) is the nominal FWER of separate
  one-sided frequentist tests at significance level \(\alpha=0.025\)
  under the global null scenario in a basket trial with \(I\) strata.
  Note that the actual FWER of the separate binomial tests may be lower
  due to the discrete nature of the binomial distribution. Furthermore,
  the TOER in every stratum in every scenario is less or equal to the
  threshold \(\eta_3 = 0.2\) from Section~\ref{sec-utility-functions}.
\item
  \emph{Moderate} (+M): The FWER in the global null scenario is less or
  equal to \(1.5\cdot \mathrm{FWER}_0\). The FWER in all scenarios is
  less or equal to \(2\cdot \mathrm{FWER}_0\). Furthermore, the TOER in
  every stratum in every scenario is less or equal to the threshold
  \(\eta_3 = 0.2\).
\item
  \emph{Liberal} (+L): The FWER in all scenarios is less or equal to
  \(3\cdot \mathrm{FWER}_0\).
\item
  \emph{Too liberal} (-L): None of the conditions above is fulfilled.
\end{itemize}

In Table~\ref{tbl-fuj03-summary}, we show an overall assessment of the
optimal tuning parameters with respect to each utility function. In the
following, we lay out a detailed justification for this assessment.

\begin{table}

\caption{\label{tbl-fuj03-summary}Assessment of optimal tuning
parameters found by the utility functions in the scenario
\texttt{sc\_fuj03}}

\centering{

\begin{tabular*}{\linewidth}{@{\extracolsep{\fill}}l|>{\raggedright\arraybackslash}p{\dimexpr 135.00pt -2\tabcolsep-1.5\arrayrulewidth}>{\raggedleft\arraybackslash}p{\dimexpr 37.50pt -2\tabcolsep-1.5\arrayrulewidth}>{\raggedleft\arraybackslash}p{\dimexpr 37.50pt -2\tabcolsep-1.5\arrayrulewidth}>{\raggedleft\arraybackslash}p{\dimexpr 37.50pt -2\tabcolsep-1.5\arrayrulewidth}}
\toprule
 &  & \multicolumn{3}{>{\centering\arraybackslash}m{\dimexpr 112.50pt -2\tabcolsep-1.5\arrayrulewidth}}{\parbox{\linewidth}{\centering {\textbf{Optimal parameters}}}} \\ 
\cmidrule(lr){3-5}
\textbf{Utility function} & \textbf{Overall assessment} & \(\lambda\) & \(\varepsilon\) & \(\tau\) \\ 
\midrule\addlinespace[2.5pt]
\texttt{fuj1} & -L & 0.99 & 2 & 0 \\ 
\texttt{fuj2} & +M & 0.99 & 2 & 0.5 \\ 
\texttt{u\_ewp} & -L & 0.99 & 2 & 0.2 \\ 
\texttt{u\_ewp\_avg} & +C & 0.99 & 25 & 0.6 \\ 
\texttt{u\_ewp\_avg\_pen} & +C & 0.99 & 25 & 0.6 \\ 
\texttt{u\_ecd} & +C & 0.99 & 25 & 0.6 \\ 
\texttt{u\_ecd\_avg} & -L & 0.99 & 2 & 0.2 \\ 
\texttt{u\_ecd\_avg\_pen} & +M & 0.999 & 2 & 0.3 \\ 
\texttt{u\_2ewp} & +C & 0.99 & 25 & 0.6 \\ 
\texttt{u\_2ewp\_avg} & +F & 0.99 & 0 & 1 \\ 
\texttt{u\_2ewp\_avg\_pen} & +F & 0.99 & 0 & 1 \\ 
\texttt{u\_2pow} & -L & 0.9 & 25 & 0.6 \\ 
\texttt{u\_2pow\_avg} & +M & 0.99 & 5 & 0 \\ 
\texttt{u\_2pow\_avg\_pen} & +M & 0.99 & 5 & 0 \\ 
\bottomrule
\end{tabular*}
\begin{minipage}{\linewidth}
\vspace{.05em}
\footnotesize\textsc{Symbols.} Overall assessment categories as defined in 
Section~\ref{sec-perf-sc-fuj03}, -C: too conservative, +F: similar to frequentist design, +C: conservative,
+M: moderate, +L: liberal, -L: too liberal. Tuning parameters, see Section~\ref{sec-basket-design}: $\lambda$ is the posterior decision threshold, $\varepsilon$ and $\tau$ determine the shape of
the borrowing weights.\\
\end{minipage}

}

\end{table}%

\begin{figure}

\centering{

\pandocbounded{\includegraphics[keepaspectratio]{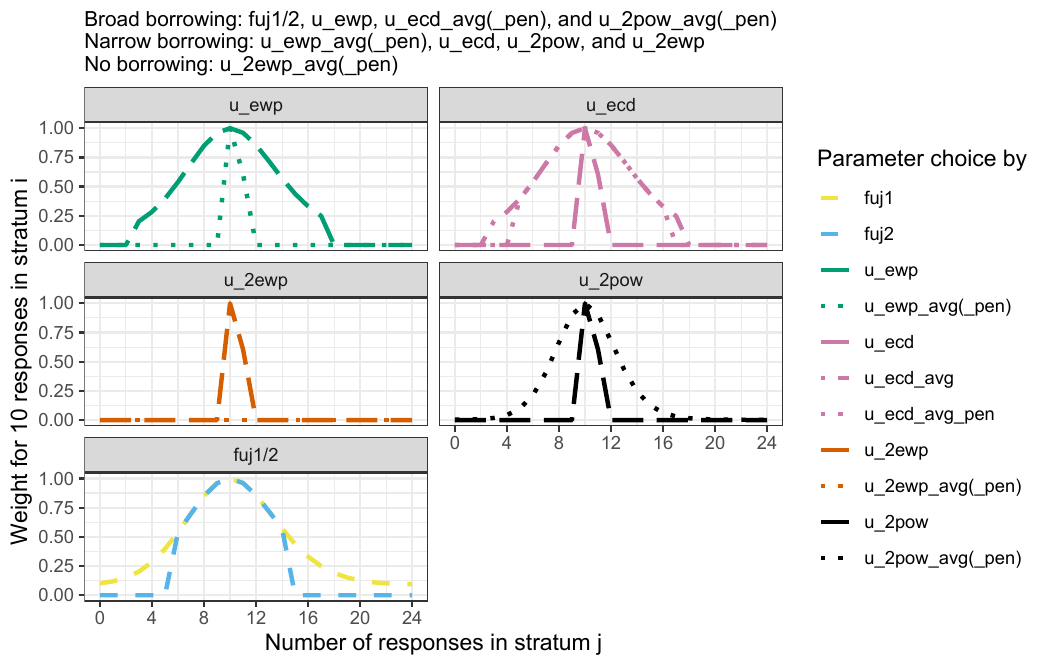}}

}

\caption{\label{fig-fuj03-weights}Weights for information borrowing in
the scenario set \texttt{sc\_fuj03}. The parameters found by
\texttt{u\_ewp\_avg} and \texttt{u\_ewp\_avg\_pen} were identical, the
same is true for \texttt{u\_2ewp\_avg(\_pen)} and
\texttt{u\_2pow\_avg(\_pen)}.\newline\footnotesize \textsc{Symbols.}
\texttt{fuj1/2}: default parameters by
\textcite{fujikawa_bayesian_2020}, \texttt{u\_*}: utility functions as
defined in Section~\ref{sec-utility-functions}.}

\end{figure}

Firstly, we will look at the optimal amount of borrowing as found by the
different utility functions. The amount of borrowing is determined by
the parameters \(\varepsilon\) and \(\tau\). Their values and the
resulting borrowing weights can be seen in
Figure~\ref{fig-fuj03-weights}. If one observes 10 responses in one
stratum \(i\), then the borrowing with another stratum \(j\) is
determined by the number of responses in that stratum. In the figure,
one sees that the optimization using \texttt{u\_ewp},
\texttt{u\_ecd\_avg(\_pen)}, and \texttt{u\_2pow\_avg(\_pen)} led to
borrowing across a broad range of response rates, as do the default
parameters \texttt{fuj1} and \texttt{fuj2} suggested by Fujikawa and
colleagues. Optimization using \texttt{u\_ewp\_avg\_pen},
\texttt{u\_ecd}, \texttt{u\_2pow}, and \texttt{u\_2ewp} led to a
borrowing in a narrow range of similar response rate estimates.
Optimization using \texttt{u\_2ewp\_avg(\_pen)} led to no borrowing at
all.

\begin{figure}

\centering{

\pandocbounded{\includegraphics[keepaspectratio]{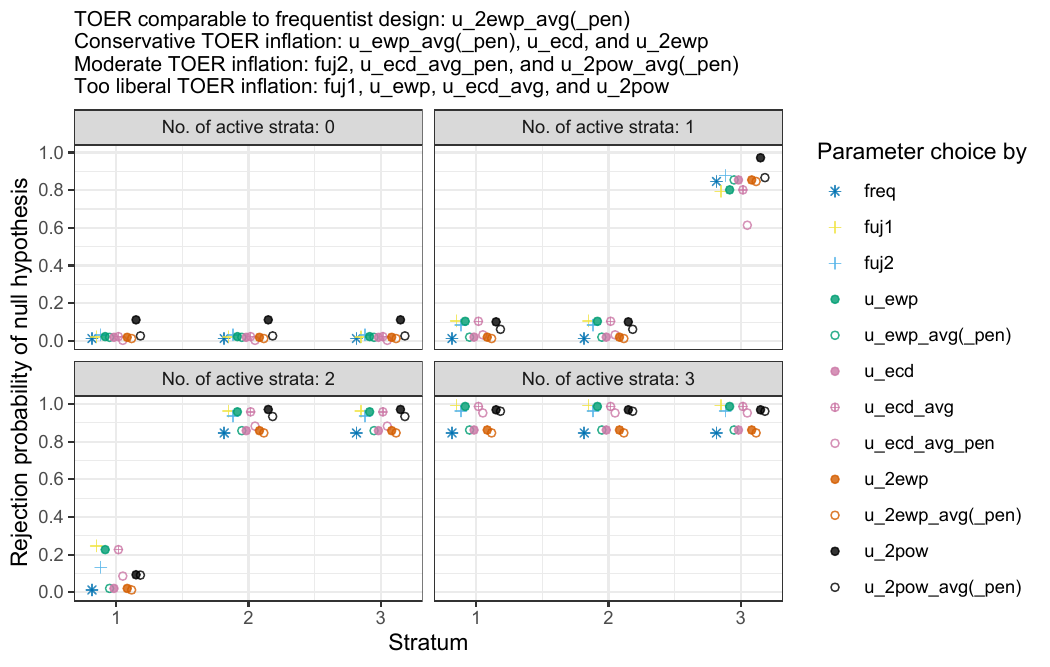}}

}

\caption{\label{fig-fuj03-all-local}Local rejection rates in the
scenario set \texttt{sc\_fuj03}. \newline\footnotesize \textsc{Symbols.}
\texttt{freq}: frequentist basket trial design with separate unadjusted
binomial tests, \texttt{fuj1/2}: default parameters by
\textcite{fujikawa_bayesian_2020}, \texttt{u\_*}: utility functions as
defined in Section~\ref{sec-utility-functions}.}

\end{figure}

As a next step, we will look at how the different amount of borrowing is
reflected in terms of local TOER and local power in each stratum as
shown in Figure~\ref{fig-fuj03-all-local}. For the default parameters
\texttt{fuj1} as well as for the optimal parameters found by
\texttt{u\_ewp} and \texttt{u\_ecd\_avg}, the borrowing range is so
broad that it leads to unfavorably high TOER in the scenario with only
one inactive stratum. Even though \texttt{u\_2pow} only borrows across a
narrow range, it shows unfavorably high local TOERs in the global null
scenario due to its more liberal decision threshold of \(\lambda = 0.9\)
compared to the most common \(0.99\). The function
\texttt{u\_ecd\_avg\_pen}, despite also showing borrowing across a broad
range, has moderate TOER inflation due to the more strict decision
threshold of \(\lambda = 0.999\). The functions \texttt{u\_2pow\_avg}
and \texttt{u\_2pow\_avg\_pen} arrived at identical parameters. They
also show borrowing across a relatively broad range of responses, but
the shape of their weights is slim enough that the TOER inflation is
moderate. Similarly, the default parameters \texttt{fuj2} also show
moderate TOER inflation. Borrowing across only a narrow range leads to
only conservative TOER inflation for the functions \texttt{u\_ewp\_avg},
\texttt{u\_ewp\_avg\_pen}, \texttt{u\_ecd}, and \texttt{u\_2ewp}.
Finally, the TOER corresponding to the parameters found by
\texttt{u\_2ewp\_avg(\_pen)} is exactly equal to the TOER of the
separate frequentist binomial tests \texttt{freq}. This is consistent
with the fact that these parameters imply no borrowing at all.

\begin{figure}

\centering{

\pandocbounded{\includegraphics[keepaspectratio]{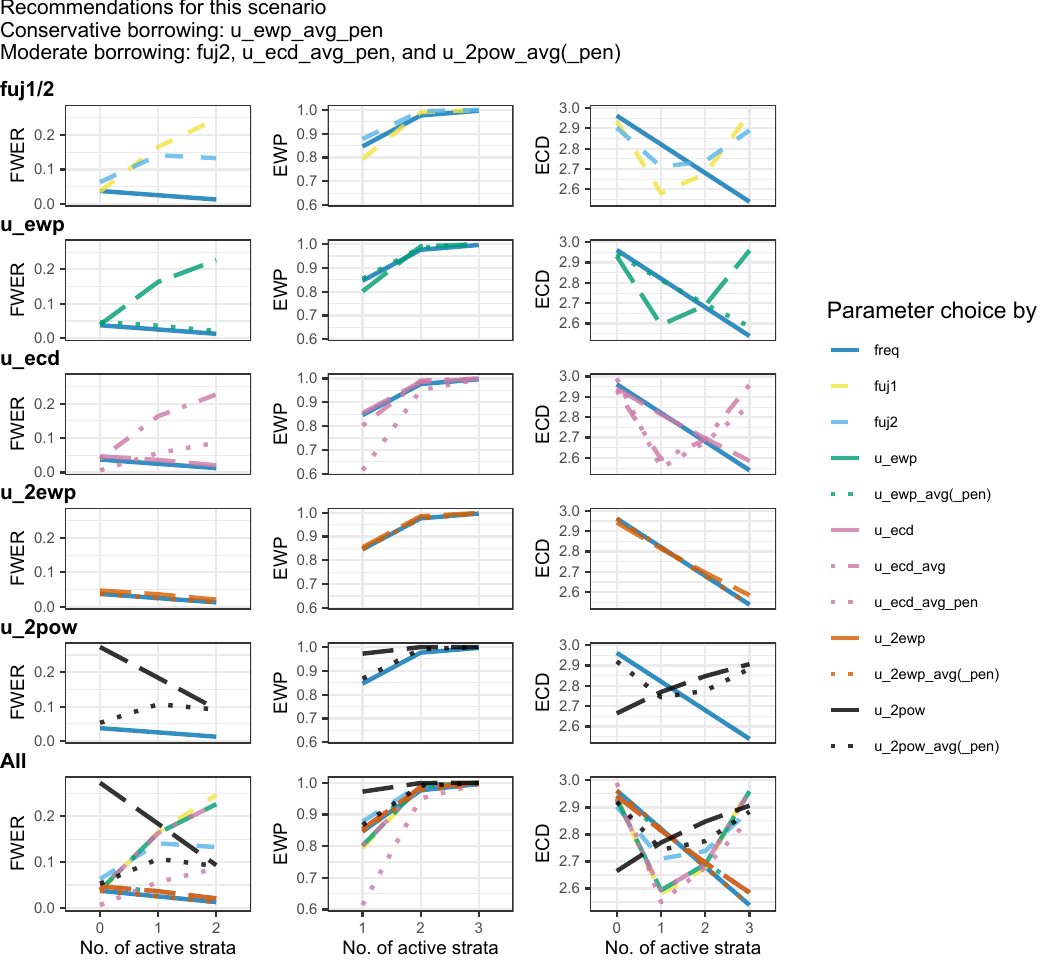}}

}

\caption{\label{fig-fuj03-all-global}Global rejection rates and correct
decisions in the scenario set \texttt{sc\_fuj03}.
\newline\footnotesize \textsc{Symbols.} FWER: family-wise error rate,
EWP: experiment-wise power, ECD: expected number of correct decisions,
\texttt{freq}: frequentist basket trial design with separate unadjusted
binomial tests, \texttt{fuj1/2}: default parameters by
\textcite{fujikawa_bayesian_2020}, \texttt{u\_*}: utility functions as
defined in Section~\ref{sec-utility-functions}.}

\end{figure}

To conclude with the performance in the scenario set \texttt{sc\_fuj03},
we look at global performance measures as shown in
Figure~\ref{fig-fuj03-all-global}. The nominal FWER for three
independent one-sided frequentist tests at level 0.025 is
\(\alpha^*=1 - (1 - 0.025)^3=0.073\). Five functions will be highlighted
here as they work well in this scenario set, and other scenario sets
described further below. Firstly, the parameters found by
\texttt{u\_ewp\_avg} respect the nominal FWER level \(\alpha^*\) while
gaining a little bit of ECD in the scenarios with a larger number of
active strata compared to the frequentist approach without borrowing.
Secondly, the parameters from \texttt{fuj2}, \texttt{u\_ecd\_avg\_pen},
\texttt{u\_2pow\_avg}, and \texttt{u\_2pow\_avg\_pen} only moderately
inflate the FWER while showing ECD gains in the scenarios with a larger
number of active strata. Taking a closer look at
\texttt{u\_ecd\_avg\_pen}, we see that the stricter decision threshold
of \(\lambda=0.999\) render the FWER almost unnecessarily conservative
for the scenarios with fewer active strata. Especially in the scenario
with only one active stratum, this results in rather low ECD. As this
function performs well in other scenario sets, we highlighted it here
nevertheless.

\subsubsection{Performance of utility functions in all scenarios --
general
remarks}\label{performance-of-utility-functions-in-all-scenarios-general-remarks}

Next we are looking at the global performance measures in all other
scenarios. We used the same overall classification of performance as in
the last section. The assessment according to this classification can be
found in Table~\ref{tbl-all-summary}. While the performance measures for
the scenario sets \texttt{sc\_fuj03} and \texttt{sc\_bau04} were
calculated exactly up to the precision of numerical integration, the
performance measures for all other scenario sets were calculated using
Monte Carlo simulations. To quantify their imprecision, Monte Carlo
standard errors (MCSEs) were calculated. All MCSEs were small, namely
below 0.01 for TOER, local power, FWER and EWP and \(0.01\cdot I\) for
the ECD. Hence, the MCSEs are considered irrelevant for the comparison
of the methods.

Concerning the results, some trends can be remarked across all scenario
sets:

The performance measure EWP is usually the least differentiated global
performance measure. This is due to its definition, being the
probability of detecting at least one truly active stratum. Especially
in scenarios with many active strata, this probability is very close to
1. In terms of performance, looking at FWER and ECD is usually
sufficient.

The performance of the parameter combination \texttt{fuj1} as well as
the performance of the parameters found by \texttt{u\_ewp},
\texttt{u2\_ewp}, \texttt{u\_2pow}, \texttt{u\_ewp\_avg},
\texttt{u\_2ewp\_avg}, \texttt{u\_ecd\_avg}, \texttt{u\_2pow\_avg}, and
\texttt{u\_2pow\_avg\_pen} had rather poor performance as explained in
the following. Hence, they are not shown in the figures in this section,
however, their performance measures can be found in the appendix.

The default parameter combination \texttt{fuj1} performs too liberally
across almost all scenario sets, see appendix,
Figure~\ref{fig-appendix-all-global-fuj12}. This is mainly due to the
fact that the decision threshold is fixed to 0.99, and, hence, FWER is
not controlled at all. Secondly, the entire family of
\texttt{u\_2ewp\_*} utility functions performed too conservatively or
similarly to the frequentist tests across all scenarios, only sometimes
showing conservative borrowing, see
Figure~\ref{fig-appendix-all-global-2ewp}. This is probably due to the
fact that the EWP is a very insensitive performance measure as discussed
above.

The single-scenario utility functions \texttt{u\_ewp} and
\texttt{u\_2pow}, which were only optimized for the scenario with
approximately half of the strata being active, often showed a too
liberal behavior in comparison to the scenario-averaged counterparts,
see Figure~\ref{fig-appendix-all-global-ewp} and
Figure~\ref{fig-appendix-all-global-2pow}. The utility function
\texttt{u\_ecd} is an exception here, delivering conservative to liberal
borrowing behavior despite only being optimized for one scenario.

The parameters found by the scenario-averaged functions were mostly very
similar or even identical to the scenario-averaged and penalized utility
functions. However, in some scenario sets, the additional penalization
resulted in more favorable performance in terms of constraining type-I
error inflation. Hence, this version is mostly preferable, see
Figure~\ref{fig-appendix-all-global-ewp},
Figure~\ref{fig-appendix-all-global-ecd}, and
Figure~\ref{fig-appendix-all-global-2pow}.

For these reasons, the performance of the aforementioned parameter
combinations is not shown in this chapter, but in the appendix. The
utility function \texttt{u\_2ewp\_avg\_pen} is shown for comparison in
the following despite its rather poor performance.

For the remaining parameter combinations, we are comparing their
performance in different groups of scenario sets, depending on whether
there are few or many strata and few or many patients.

\begin{table}

\caption{\label{tbl-all-summary}Assessment of optimal tuning parameters
found by the utility functions in all scenarios}

\centering{

\begin{tabular*}{\linewidth}{@{\extracolsep{\fill}}l|>{\raggedright\arraybackslash}p{\dimexpr 45.00pt -2\tabcolsep-1.5\arrayrulewidth}>{\raggedright\arraybackslash}p{\dimexpr 45.00pt -2\tabcolsep-1.5\arrayrulewidth}>{\raggedright\arraybackslash}p{\dimexpr 45.00pt -2\tabcolsep-1.5\arrayrulewidth}>{\raggedright\arraybackslash}p{\dimexpr 45.00pt -2\tabcolsep-1.5\arrayrulewidth}>{\raggedright\arraybackslash}p{\dimexpr 45.00pt -2\tabcolsep-1.5\arrayrulewidth}>{\raggedright\arraybackslash}p{\dimexpr 45.00pt -2\tabcolsep-1.5\arrayrulewidth}>{\raggedright\arraybackslash}p{\dimexpr 45.00pt -2\tabcolsep-1.5\arrayrulewidth}}
\toprule
 & \multicolumn{7}{>{\centering\arraybackslash}m{\dimexpr 315.00pt -2\tabcolsep-1.5\arrayrulewidth}}{\parbox{\linewidth}{\centering {\textbf{Scenarios with}}}} \\ 
\cmidrule(lr){2-8}
 & \multicolumn{3}{>{\centering\arraybackslash}m{\dimexpr 135.00pt -2\tabcolsep-1.5\arrayrulewidth}}{\parbox{\linewidth}{\centering {\textbf{few strata, few patients}}}} & \multicolumn{1}{>{\centering\arraybackslash}m{\dimexpr 45.00pt -2\tabcolsep-1.5\arrayrulewidth}}{\parbox{\linewidth}{\centering {\textbf{few strata, many pat.s}}}} & \multicolumn{1}{>{\centering\arraybackslash}m{\dimexpr 45.00pt -2\tabcolsep-1.5\arrayrulewidth}}{\parbox{\linewidth}{\centering {\textbf{many strata, few pat.s}}}} & \multicolumn{2}{>{\centering\arraybackslash}m{\dimexpr 90.00pt -2\tabcolsep-1.5\arrayrulewidth}}{\parbox{\linewidth}{\centering {\textbf{many strata, many patients}}}} \\ 
\cmidrule(lr){2-4} \cmidrule(lr){5-5} \cmidrule(lr){6-6} \cmidrule(lr){7-8}
\textbf{Parameter choice by} & sc\_fuj03 & sc\_bau04 & sc\_med04 & sc\_lrg03 & sc\_kra08 & sc\_med09 & sc\_lrg20 \\ 
\midrule\addlinespace[2.5pt]
\texttt{fuj1} & -L & -L & -L & -L & -L & -L & +L \\ 
\texttt{fuj2} & +M & +L & +M & +L & -L & -L & +L \\ 
\texttt{u\_ewp} & -L & -L & -L & +M & -L & -C & -C \\ 
\texttt{u\_ewp\_avg} & +C & -L & +C & +C & +M & -C & -C \\ 
\texttt{u\_ewp\_avg\_pen} & +C & +M & +C & +C & +M & -C & -C \\ 
\texttt{u\_ecd} & +C & +M & +C & +L & +M & -C & -C \\ 
\texttt{u\_ecd\_avg} & -L & -L & +M & -L & +M & -C & -C \\ 
\texttt{u\_ecd\_avg\_pen} & +M & +L & +M & +M & +M & -C & -C \\ 
\texttt{u\_2ewp} & +C & +F & -C & +C & +F & +F & -C \\ 
\texttt{u\_2ewp\_avg} & +F & +F & -C & +C & +F & -C & -C \\ 
\texttt{u\_2ewp\_avg\_pen} & +F & +F & -C & +C & +F & -C & -C \\ 
\texttt{u\_2pow} & -L & +L & +F & -L & +M & -L & +M \\ 
\texttt{u\_2pow\_avg} & +M & +L & +F & -L & +L & +F & +M \\ 
\texttt{u\_2pow\_avg\_pen} & +M & +L & +F & -L & +L & +F & +M \\ 
\bottomrule
\end{tabular*}
\begin{minipage}{\linewidth}
\vspace{.05em}
\footnotesize\textsc{Symbols.} Overall assessment categories, see 
Section~\ref{sec-perf-sc-fuj03}, -C: too conservative, +F: similar to frequentist design, +C: conservative,
+M: moderate, +L: liberal, -L: too liberal.\\
\end{minipage}

}

\end{table}%

\subsubsection{Performance of utility functions in scenarios with few
strata and few
patients}\label{performance-of-utility-functions-in-scenarios-with-few-strata-and-few-patients}

\begin{figure}

\centering{

\pandocbounded{\includegraphics[keepaspectratio]{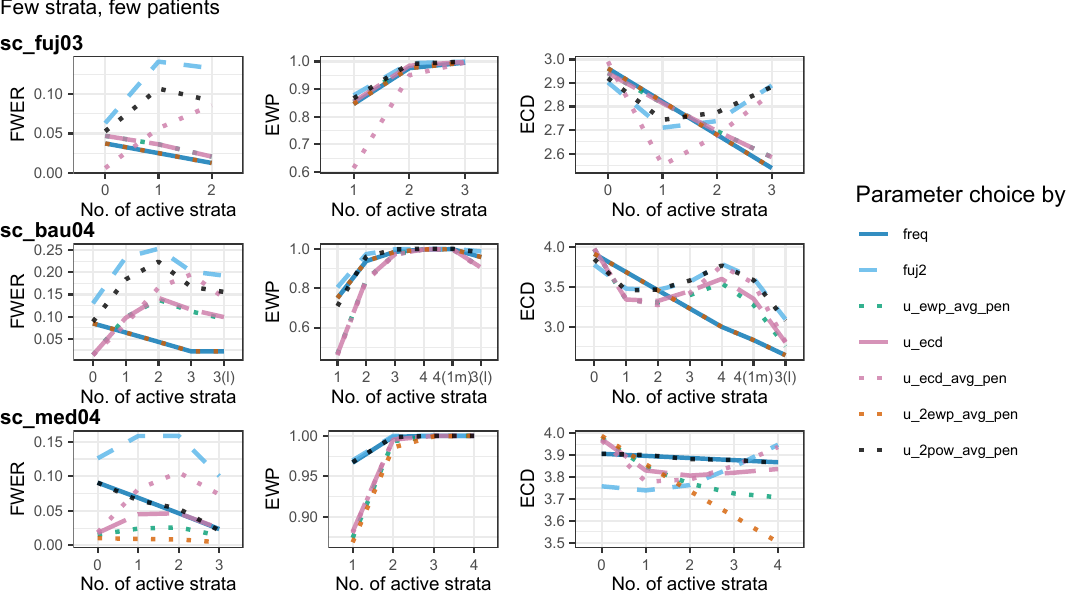}}

}

\caption{\label{fig-fewstratafewpat-all-global}Global rejection rates
and correct decisions in the scenario sets with few strata and few
patients, \texttt{sc\_fuj03}, \texttt{sc\_bau04}, \texttt{sc\_med04}.
\newline\footnotesize \textsc{Symbols.} 4(1m): scenario with 4 active
strata and `one in the middle', 3(l): scenario with 3 active strata and
`linear' increase in response, see Table~\ref{tbl-scenario-sets}. FWER:
family-wise error rate, EWP: experiment-wise power, ECD: expected number
of correct decisions, \texttt{freq}: frequentist basket trial design
with separate unadjusted binomial tests, \texttt{fuj1/2}: default
parameters by \textcite{fujikawa_bayesian_2020}, \texttt{u\_*}: utility
functions as defined in Section~\ref{sec-utility-functions}.}

\end{figure}

We start by looking at the global performance in the scenario sets with
few strata and few patients as shown in
Figure~\ref{fig-fewstratafewpat-all-global}, namely the scenarios sets
\texttt{sc\_fuj03}, \texttt{sc\_bau04}, and \texttt{sc\_med04}. One
thing that quickly jumps to the eye is the fact that, almost
homogeneously across all utility functions, the performance in the
\texttt{sc\_bau04} scenario set is slightly more liberal than in the
scenario sets \texttt{sc\_fuj03} and \texttt{sc\_med04}. This is
presumably due to the fact that this scenario set contains two more
non-null scenarios (`one in the middle' and `linear'), resulting in more
weight on the non-null scenarios. The utility functions \texttt{u\_ecd}
and \texttt{u\_ewp\_avg\_pen} resulted in conservative to moderate
borrowing, with the FWER being still fairly close to the FWER of the
frequentist design while the ECD is increased in the scenarios with
greater numbers of active strata in the scenario sets \texttt{sc\_fuj03}
and \texttt{sc\_med04}. Note that in the scenario set
\texttt{sc\_med04}, the ECD is worse than the frequentist design's for
these two utility functions. These functions have a fixed FWER
constraint to 0.05. Due to the slightly larger sample size in this
scenario set, this threshold is crossed by the one-sided frequentist
tests. As mentioned before, the utility function
\texttt{u\_2ewp\_avg\_pen} is too conservative or identical to the
frequentist test. The parameters \texttt{fuj2} as well as those found by
\texttt{u\_2pow\_avg\_pen} and \texttt{u\_ecd\_avg\_pen} mostly show
moderate to liberal borrowing. This can be seen clearly in the increase
in FWER with increasing number of active strata, which coincides with a
gain in ECD in scenarios with many active strata. Of these three utility
functions, the \texttt{u\_ecd\_avg\_pen} function has the lowest FWER
due to its built-in constraint of FWER under the global null. This
performance is achieved by keeping moderate borrowing in the parameters
\((\varepsilon, \tau)\) while raising the posterior decision threshold
to \(\lambda = 0.999\). The function \texttt{u\_2pow\_avg\_pen} on the
other hand restricts to no borrowing in the scenario \texttt{sc\_med04},
again probably due to the slightly larger sample size. To summarize, one
can say that information borrowing in basket trial designs with few
strata and few patients can be optimized using \texttt{u\_ecd} or
\texttt{u\_ewp\_avg\_pen} if one wants conservative borrowing, and using
\texttt{fuj2}, \texttt{u\_ecd\_avg\_pen}, or \texttt{u\_2pow\_avg\_pen}
if one wants moderate to liberal borrowing.

\subsubsection{Performance of utility functions in scenarios with few
strata and many
patients}\label{performance-of-utility-functions-in-scenarios-with-few-strata-and-many-patients}

\begin{figure}

\centering{

\pandocbounded{\includegraphics[keepaspectratio]{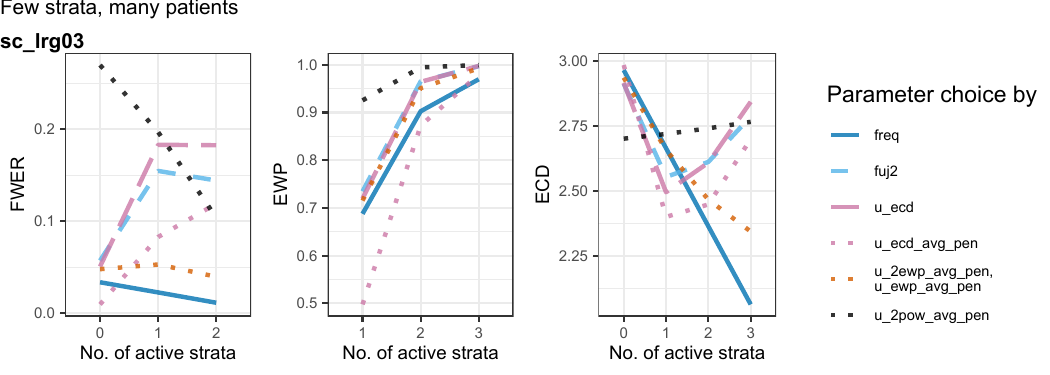}}

}

\caption{\label{fig-fewstratamanypat-all-global}Global rejection rates
and correct decisions in the scenario set with few strata and many
patients, \texttt{sc\_lrg03}. \texttt{u\_ewp\_avg\_pen} and
\texttt{u\_2ewp\_avg\_pen} had identical optimal parameters.
\newline\footnotesize  \textsc{Symbols.} FWER: family-wise error rate,
EWP: experiment-wise power, ECD: expected number of correct decisions,
\texttt{freq}: frequentist basket trial design with separate unadjusted
binomial tests, \texttt{fuj1/2}: default parameters by
\textcite{fujikawa_bayesian_2020}, \texttt{u\_*}: utility functions as
defined in Section~\ref{sec-utility-functions}.}

\end{figure}

Next, let us look at the global performance measures in the scenario set
with few strata but many patients, \texttt{sc\_lrg03}, as shown in
Figure~\ref{fig-fewstratamanypat-all-global}. The utility functions
\texttt{u\_ewp\_avg\_pen} and \texttt{u\_2ewp\_avg\_pen} show identical
conservative borrowing with only a little FWER inflation. The utility
function \texttt{u\_ecd\_avg\_pen} shows moderate borrowing. The default
parameters \texttt{fuj2} as well as the parameters found by
\texttt{u\_ecd} show liberal borrowing similar to the scenario, with
performance patterns similar to the scenario sets with few strata and
few patients. Contrary to that, the performance of the function
\texttt{u\_2pow\_avg\_pen} is quite different. While the ECD is higher
than the frequentist test's even in the scenario with one active
stratum, this advantage is bought with too liberal FWER inflation under
the global null hypothesis. To summarize, information borrowing in
basket trial designs with few strata and many patients can be optimized
using \texttt{u\_ewp\_avg\_pen} and \texttt{u\_2ewp\_avg\_pen} if
conservative borrowing is desired, using \texttt{u\_ecd\_avg\_pen} if
moderate borrowing is desired, and \texttt{fuj2} or \texttt{u\_ecd} if
liberal borrowing is desired.

\subsubsection{Performance of utility functions in scenario with many
strata and few
patients}\label{performance-of-utility-functions-in-scenario-with-many-strata-and-few-patients}

\begin{figure}

\centering{

\pandocbounded{\includegraphics[keepaspectratio]{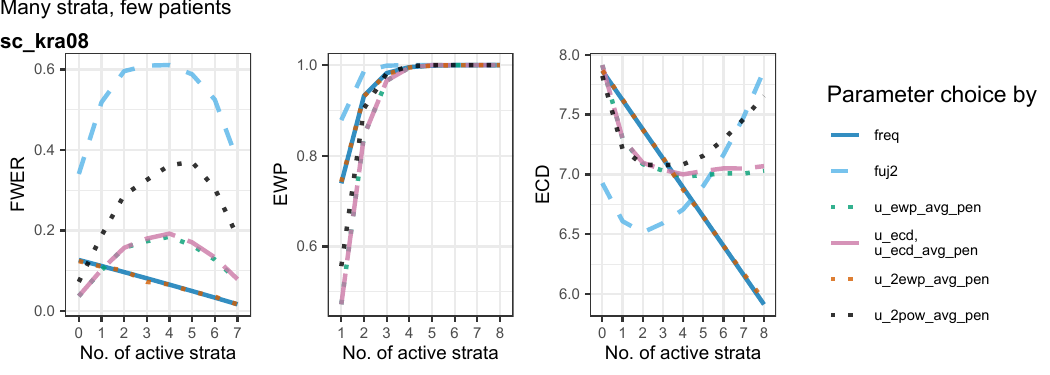}}

}

\caption{\label{fig-manystratafewpat-all-global}Global rejection rates
and correct decisions in the scenario sets with many strata and few
patients. \texttt{u\_ecd} and \texttt{u\_ecd\_avg\_pen} had identical
optimal parameters. \newline\footnotesize  \textsc{Symbols.} FWER:
family-wise error rate, EWP: experiment-wise power, ECD: expected number
of correct decisions, \texttt{freq}: frequentist basket trial design
with separate unadjusted binomial tests, \texttt{fuj1/2}: default
parameters by \textcite{fujikawa_bayesian_2020}, \texttt{u\_*}: utility
functions as defined in Section~\ref{sec-utility-functions}.}

\end{figure}

Now, we look at the performance of the utility functions in the scenario
set with many strata and few patients, \texttt{sc\_kra08}, as shown in
Figure~\ref{fig-manystratafewpat-all-global}. The default parameters
\texttt{fuj2} perform too liberally, whereas the utility function
\texttt{u\_2ewp\_avg\_pen} is almost identical to the separate
frequentist tests (the minor deviations may be due to random imprecision
of the Monte Carle procedure). The utility functions \texttt{u\_ecd},
\texttt{u\_ecd\_avg\_pen} and \texttt{u\_ewp\_avg\_pen} all delivered
moderate borrowing with some FWER inflation and gains in ECD for
scenarios with many active strata. The utility function
\texttt{u\_2pow\_avg\_pen} resulted in liberal borrowing behavior with
even higher FWER inflation and higher ECD gains for the scenarios with
many active strata. In summary, in the scenario with many strata and few
patients, optimizing with \texttt{u\_ecd}, \texttt{u\_ecd\_avg\_pen} and
\texttt{u\_ewp\_avg\_pen} leads to moderate borrowing, optimizing with
\texttt{u\_2pow\_avg\_pen} leads to liberal borrowing.

\subsubsection{Performance of utility functions in scenarios with many
strata and many
patients}\label{performance-of-utility-functions-in-scenarios-with-many-strata-and-many-patients}

\begin{figure}

\centering{

\pandocbounded{\includegraphics[keepaspectratio]{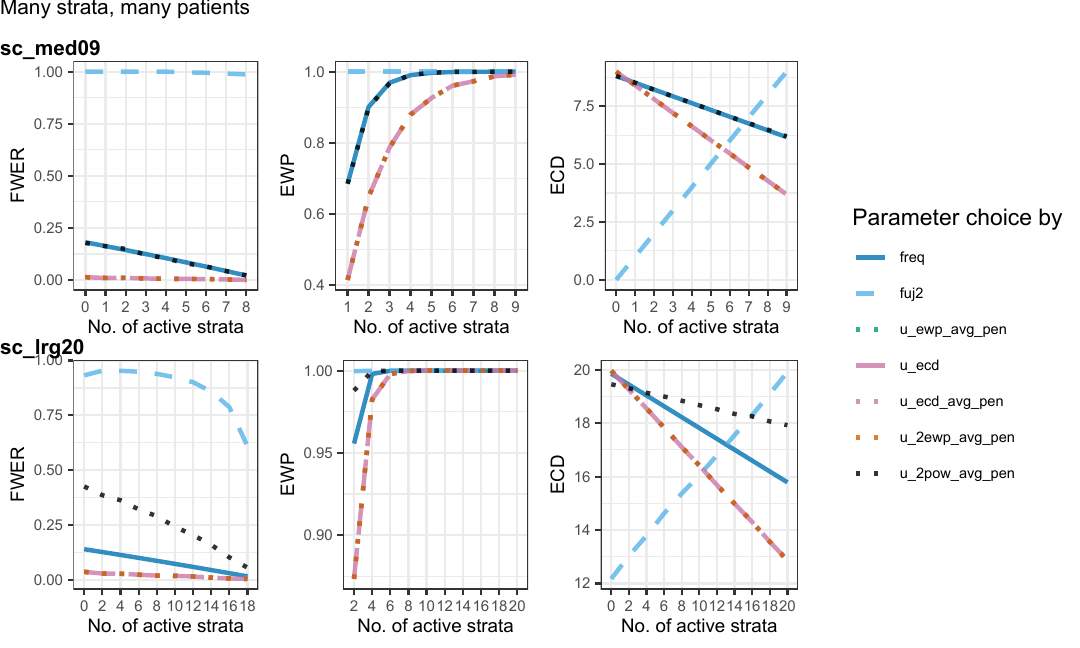}}

}

\caption{\label{fig-manystratamanypat-all-global}Global rejection rates
and correct decisions in the scenario sets with many strata and many
patients. The performance of \texttt{u\_ecd}, \texttt{u\_ecd\_avg\_pen},
\texttt{u\_2ewp\_avg\_pen} and \texttt{u\_ewp\_avg\_pen} is exactly
identical, hence not all lines can be seen.
\newline\footnotesize  \textsc{Symbols.} FWER: family-wise error rate,
EWP: experiment-wise power, ECD: expected number of correct decisions,
\texttt{freq}: frequentist basket trial design with separate unadjusted
binomial tests, \texttt{fuj1/2}: default parameters by
\textcite{fujikawa_bayesian_2020}, \texttt{u\_*}: utility functions as
defined in Section~\ref{sec-utility-functions}.}

\end{figure}

Finally, let us look at the performance of the utility functions in the
scenario sets with many strata and many patients, \texttt{sc\_med09} and
\texttt{sc\_lrg20}, which are shown in
Figure~\ref{fig-manystratamanypat-all-global}. The default parameters
\texttt{fuj2} fail in these scenario sets. This is due to a combination
of two things: The fact that the design by Fujikawa et al.~does not
limit the amount of borrowing and the fact that the posterior decision
threshold \(\lambda\) is fixed to 0.99 imply that posterior information
of the arm itself can be ``overwhelmed'' by the information from the
other arms. Altering Fujikawa's design by restricting the amount of
information borrowed from other arms would be helpful here; however, the
effect would be similar to reducing the amount of borrowing altogether.
All utility functions arrived at parameters resulting in no borrowing at
all, so the differences in FWER and ECD are only due to the decision
threshold \(\lambda\). The utility functions \texttt{u\_ecd},
\texttt{u\_ecd\_avg\_pen}, \texttt{u\_ewp\_avg\_pen} and
\texttt{u\_2ewp\_avg\_pen} showed FWER lower than the frequentist level,
\texttt{u\_2pow\_avg\_pen} showed FWER greater or equal to the
frequentist level. In summary, no utility function found any borrowing
to be optimal and the performance characteristics can be tuned by the
decision threshold only.

\subsubsection{Performance of utility functions in all scenarios --
summary}\label{performance-of-utility-functions-in-all-scenarios-summary}

To summarize, we saw that the utility functions under consideration
resulted in different performance in terms of FWER inflation and ECD
gains. The scenario-averaged and penalized utility functions usually
showed a more robust compromise of performance characteristics, with
\texttt{u\_ewp\_avg\_pen} resulting in mostly conservative,
\texttt{u\_ecd\_avg\_pen} resulting in mostly moderate, and
\texttt{u\_2pow\_avg\_pen} resulting in moderate to liberal information
borrowing. The single-scenario function \texttt{u\_ecd} performed well,
but its behavior was slightly more erratic, as the function resulted in
conservative to liberal information borrowing parameters across the
scenario sets. The default parameters \texttt{fuj2} with
\((\lambda,\varepsilon,\tau)=(0.99, 2, 0.5)\) also performed fairly
well, but it is recommended to tune the decision threshold \(\lambda\)
in order to ensure that FWER inflation does not exceed an acceptable
threshold. In any case, information borrowing resulted in type-I error
inflation in some scenarios. In order to ensure type-I error control in
all scenarios, frequentist testing would be the safest option. A brief
summary of the recommendations can be found in
Table~\ref{tbl-part2-recommendations}.

\begin{table}

\caption{\label{tbl-part2-recommendations}Recommendations based on the
comparison study results}

\centering{

\begin{tabular*}{\linewidth}{@{\extracolsep{\fill}}>{\raggedright\arraybackslash}p{\dimexpr 82.50pt -2\tabcolsep-1.5\arrayrulewidth}>{\raggedright\arraybackslash}p{\dimexpr 75.00pt -2\tabcolsep-1.5\arrayrulewidth}>{\raggedright\arraybackslash}p{\dimexpr 67.50pt -2\tabcolsep-1.5\arrayrulewidth}>{\raggedright\arraybackslash}p{\dimexpr 180.00pt -2\tabcolsep-1.5\arrayrulewidth}}
\toprule
\textbf{Trial setting} & \textbf{Trial setting specification} & \textbf{Scenario sets} & \textbf{Optimized information borrowing} \\ 
\midrule\addlinespace[2.5pt]
\multicolumn{4}{>{\raggedright\arraybackslash}m{405pt}}{\parbox{\linewidth}{Analysis strategy: Local type-I error control in every stratum}} \\[2.5pt] 
\midrule\addlinespace[2.5pt]
Any trial setting &  &  & do not borrow, use frequentist testing procedure \\ 
\midrule\addlinespace[2.5pt]
\multicolumn{4}{>{\raggedright\arraybackslash}m{405pt}}{\parbox{\linewidth}{Analysis strategy: Optimized information borrowing}} \\[2.5pt] 
\midrule\addlinespace[2.5pt]
Few strata, few patients & \(I \leq 4\) and \(n_{\text{total}} \leq 150\) & \texttt{sc\_fuj03}, \texttt{sc\_bau04}, \texttt{sc\_med04} & conservative to moderate: \texttt{u\_ewp\_avg\_pen},

moderate to liberal: \texttt{fuj2}, \texttt{u\_ecd\_avg\_pen} and \texttt{u\_2pow\_avg\_pen} \\ 
Few strata, many patients & \(I \leq 4\) and \(n_{\text{total}} > 150\) & \texttt{sc\_lrg03} & conservative: \texttt{u\_ewp\_avg\_pen}

moderate: \texttt{u\_ecd\_avg\_pen},

liberal: \texttt{fuj2} \\ 
Many strata, few patients & \(I \geq 8\) and \(n_{\text{total}} \leq 200\) & \texttt{sc\_kra08} & moderate: \texttt{u\_ewp\_avg\_pen}, \texttt{u\_ecd\_avg\_pen},

liberal: \texttt{u\_2pow\_avg\_pen} \\ 
Many strata, many patients & \(I \geq 8\) and \(n_{\text{total}} > 200\) & \texttt{sc\_med09}, \texttt{sc\_lrg20} & do not borrow, tune power and type-I error solely with the decision threshold \\ 
\bottomrule
\end{tabular*}
\begin{minipage}{\linewidth}
\vspace{.05em}
\footnotesize\textsc{Borrowing specification.}  Conservative: FWER inflation no more than 50\% of frequentist global null FWER in all scenarios, local TOER below 0.2 in all strata and scenarios. Moderate: FWER inflation no more than 100\% of frequentist global null FWER in all scenarios, local TOER below 0.2 in all strata and scenarios. Liberal: FWER inflation no more than 200\% of frequentist global null FWER, no further constraints, see 
Section~\ref{sec-perf-sc-fuj03}.\\
\end{minipage}

}

\end{table}%

\section{Discussion}\label{sec-discussion}

In this article, we provided a framework which attempts to address all
aspects relevant to planning the statistical aspects of a basket trial
design in a concise checklist. As type-I error and power remain central
measures of validity and efficiency in clinical trials, the following
question arises: Are power gains possible in basket trials when strict
type-I error control is required? In the context of information
borrowing from historical data, this is not possible due to the
existence of a UMP test in many common situations
\autocite{kopp-schneider_power_2020}. This is different in the context
of basket trials. In the appendix, Section~\ref{sec-counterex}, we
present a mathematical counterexample to demonstrate the following
proposition: \emph{In general, no UMP test exists in the class of level
\(\alpha\) tests for all one-sided alternative hypotheses in a basket
trial design.}

This fact can be seen as both a warning and a justification for the
development of information borrowing methods for basket trials: It is
mathematically impossible that an information borrowing design becomes a
``jack of all trades'' which achieves the highest power in all
alternative scenarios while respecting type-I error in all possible null
hypotheses. Nevertheless, for some alternative hypotheses, information
borrowing may be better compared to the classical frequentist methods,
and can even be tuned to respect type-I error constraints for select
null hypotheses. One should keep in mind though that most information
borrowing designs were not primarily designed for type-I error control
and type-I error is often calculated based on simulations. For strict
type-I error control, it may be more sensible to apply a multiple
testing procedure, e.g.~based on the closed testing principle. In any
case, it is essential to tune basket trial designs to optimal
performance in a set of pre-specified scenarios and to communicate the
benefits and shortcomings of this optimal design. A scoping review on
information borrowing using external data or internal data from other
arms observed that many trials did not provide information on how the
parameters influencing the degree of borrowing were chosen
\autocite{weru_uptake_2026}. A clear specification of the optimization
procedure would be useful whenever the tuning parameters are to be
chosen based on the trial's operating characteristics.

Two essential issues in optimizing information borrowing are the choice
of a reliable and efficient optimization procedure and the choice of a
good target for optimization. As mathematical solutions for optimality
are often hard to obtain, we addressed the former issue with choosing
optimization algorithms and the latter issue with defining different
utility functions.

In part I of our comparison study, we saw that grid search was both the
fastest and the most reliable optimization algorithm. One should note,
however, that the speed of grid search benefited from two things:
Firstly, grid search was parallelized on 50 cores. All other
optimization algorithms were sequential and could hence not be
parallelized. If fewer cores had been available, this advantage in speed
would have been less pronounced. In order to speed up sequential
algorithms, one can implement an early stopping rule if only very small
increments in the utility function are achieved. This decreases run time
of sequential algorithms and usually only negligibly affects the
optimality of the result. While this kind of early stopping rule was
used in the run of COBYLA, it was not implemented for the other
sequential optimization algorithms. Secondly, grid search benefited from
the relatively low number of three tuning parameters
\((\lambda, \varepsilon, \tau)\). As the run time of grid search grows
exponentially in the number of parameters, a sequential algorithm might
perform faster in basket trial designs with many tuning parameters. To
summarize, the results of part I demonstrate an advantage of the ``brute
force'' grid search algorithm, which might have been less pronounced
given a different basket trial design, different computational resources
or more sophisticated adjustments to the competing sequential
algorithms. Grid search has already been in use in the optimization of
basket trial designs
\autocite{jiang_optimal_2021,broglio_comparison_2022,baumann_basket_2024a},
presumably because of its simple implementation.

In part II, we saw that different utility functions achieved different
compromises in terms of frequentist performance measures FWER and ECD.
Roughly speaking, one can say the utility functions
\texttt{u\_ewp\_avg\_pen}, \texttt{u\_ecd\_avg\_pen} and
\texttt{u\_2pow\_avg\_pen} mostly achieved conservative, moderate and
liberal borrowing behavior, respectively. Nevertheless, their
performance depended on the number of strata and the number of patients
as laid out in the recommendations in
Table~\ref{tbl-part2-recommendations}. The results underline that tuning
the decision threshold with respect to the relevant outcome scenario set
is essential in achieving a desired level of type-I error control. The
utility functions \texttt{u\_ecd\_avg\_pen} and
\texttt{u\_2pow\_avg\_pen} are penalized versions of utility functions
used in the literature already
\autocite{jiang_optimal_2021,broglio_comparison_2022}. Furthermore, the
results showed that borrowing becomes less beneficial in scenarios with
increasing number strata and patients. Fujikawa's basket trial design,
which we considered in this comparison study, is especially sensitive to
this tendency as it does not restrict the amount of information
borrowed. Hence, a stratum may easily be ``overwhelmed'' if it borrows
information from other strata. A modified design which limits the
maximal amount of borrowing might still allow for some borrowing. One
approach to limit the amount of borrowing involves cutting of weights
depending on the sample sizes in the two strata, which has been
suggested for basket trials with unequal sample sizes
\autocite{schmitt_systematic_2026}. Another idea would be to estimate
the effective sample size (ESS) as suggested by
\autocite{chen_enhancing_2025} as a measure of the amount of information
borrowed. This ESS could then be limited in the tuning process of the
design.

Our results concerning the performance of the different utility
functions should be interpreted carefully, as the performance also
depends on the parameters of the utility functions. For example, the
FWER threshold in the definition of the \texttt{u\_ewp\_*} and
\texttt{u\_ecd\_*} families was set to 0.05. For basket trial design
with \(I=2\) strata, this is close to the nominal FWER of separate
one-sided frequentist tests, \(1-(1-0.025)^2=0.049\). To deepen
understanding of the utility functions, future research should look at
varying the parameters of the utility functions with respect to the
nominal FWERs of the scenario at hand. For trials with many strata,
higher FWER may be deemed acceptable compared to trials with few strata.

Another limitation of our results is that only one type of basket trial
design was considered: The basket trial design by
\textcite{fujikawa_bayesian_2020} is attractive because of the simple
form of its posterior distribution which allows for exact calculation of
performance characteristics in settings with few strata and patients. It
is also untypical in the respect that borrowing is implemented by
empirically altering the posterior. Contrary to that, basket trial
designs based on the Bayesian hierarchical model
\autocites[BHM,][]{thall_hierarchical_2003,berry_bayesian_2013} and its
extensions such as the exchangeability non-exchangeability design
\autocite[EXNEX,][]{neuenschwander_robust_2016} allow for information
borrowing by using a hyperprior which models the underlying parameters
of the prior distributions in all strata. Future work on the
optimization of basket trial designs involves application of the
framework to other basket trial designs such as the BHM and EXNEX. As
the suggested framework is independent of the choice of information
borrowing mechanism, it is also applicable to such designs. An
interesting competitor to the presented framework is the approach by
\textcite{maulik_bayesian_2026}. There, information borrowing does not
take place at the modelling stage. Instead a loss function combines the
test decisions of separate models in each strata. A Bayesian decision
rule is then derived which minimizes the posterior expected loss.

Furthermore, the comparison study was mostly limited to scenario sets in
which there is one fixed response rate for inactive strata and another
fixed response rate for active strata. Only the scenario set
\texttt{sc\_bau04} involved more complex response scenarios such as a
``linear increase'' in response rate in the different strata and an
alternative scenario ``one in the middle'' with all strata being active
but at different response rates. The restriction to scenario sets with
two fixed response rates allows for clearly recognizing trends in global
performance measures. However in actual clinical trials, the relevant
response scenarios would need to be investigated on a case by case
basis.

Finally, the recommendations divided by ``few strata'' vs.~``many
strata'' are based on scenario sets with \(I\leq 4\) and \(I\geq 8\).
More research is needed to see how the utility functions considered here
perform in scenario sets with \(I=5,6,7\) strata.

Despite these limitations in terms of utility parameters, choice of
information borrowing mechanism and choice of response scenarios, the
results of our comparison study provide valuable insights concerning the
choice of an appropriate tuning target when optimizing basket trial
designs. It shows that the precise specification of a utility function
may result in very different global performance measures. In particular,
communication of the performance measures in all relevant scenarios is
crucial to understanding the effect of information borrowing.

As an outlook, we would like to point out that the framework presented
was only applied to one-stage basket trial designs. Basket trial designs
involving two or more analysis stages with early stops for futility or
efficacy have been suggested \autocite{pohl_categories_2021}, sample
size recalculation could also be considered. The literature on
group-sequential designs also implicitly or explicitly involves the
selection of utility functions targeted in an optimization procedure.
Among these utility functions are the expected sample size under the
null hypothesis \(H_0\) and the maximal sample size
\autocite{simon_optimal_1989}, a weighted combination of the expected
sample sizes under \(H_0\) and an alternative hypothesis \(H_1\)
\autocite{pilz_optimal_2021}, and the probability of early termination
under \(H_0\) \autocite{freitag_optimal_2024}. For an optimal
group-sequential basket trial design the utility functions for
optimizing information borrowing and the utility functions for
optimizing the group-sequential design elements would need to be
combined. Practical interpretability of these combined utility functions
could be achieved by incorporating quantitative estimates of the cost
and benefit of incorrectly/correctly concluding that a drug is effective
for an active/inactive disease stratum. This idea has been discussed by
\textcite{jiang_optimal_2021} in the context of basket trials and by
\textcite{maurer_optimal_2023} in the context of multiple test
procedures.

\section{Conclusion}\label{sec-conclusion}

In conclusion, this article intends to provide guidance in optimizing
information borrowing in basket trial designs independent of the
specific choice of the statistical method. An abundance of statistical
methods for information borrowing have been suggested in the literature
as summarized by \autocite{pohl_categories_2021} and yet more designs
have been suggested since 2021. However after optimization, different
statistical methods for information borrowing have shown rather similar
performance \autocite{broglio_comparison_2022,baumann_basket_2024a}.
Despite this large number of suggested methods, the uptake in actual
clinical trials has so far been slow. Surely, the reasons for this are
manifold. However, given the rather small differences in performance,
optimization and clear communication of design choices may be just as
relevant for the uptake of information borrowing methods as the
development of new variants of such methods. This is further underlined
by the proposition presented above which suggests that no design is
optimal in every trial setting. The framework and the results of our
comparison study may aid in communicating the effect of an optimized
basket trial design on the trial's performance characteristics.

\section*{Data availability}\label{data-availability}
\addcontentsline{toc}{section}{Data availability}

\small

The program code and the output data sets generated and analysed during
the comparison study are available on GitHub under
\url{https://github.com/LukasDSauer/utility-based-optimization-of-basket-trials}.

\section*{Acknowledgements}\label{acknowledgements}
\addcontentsline{toc}{section}{Acknowledgements}

We gratefully acknowledge funding by the German Research Foundation (see
Section ``Funding'') as well computational support by means of the
high-performance computing cluster \emph{BinAC2} of the University of
Tübingen, Germany, which is part of the \emph{bwHPC} network of the
German state of Baden-Württemberg. LDS expresses his gratitude to his
thesis advisory committee members, Carolin Herrmann, Heinrich Heine
University Düsseldorf, Germany, and Norbert Benda, University Medical
Center Göttingen, Germany, for valuable comments and critique throughout
the course of the project. Furthermore, he thanks numerous colleagues at
the Institute of Medical Biometry, Heidelberg, Germany, at the German
Cancer Research Center, Heidelberg, Germany, and at the Charité Berlin,
Germany, for valuable discussions and support.

\section*{Funding}\label{funding}
\addcontentsline{toc}{section}{Funding}

This project was funded by the German Research Organization
(\emph{Deutsche Forschungsgemeinschaft}, DFG) as part of the project
\emph{STOP OR GO}, grant KI 708/9-1.

\normalsize

\section*{References}\label{references}
\addcontentsline{toc}{section}{References}

\printbibliography[heading=none]

\newpage
\setcounter{section}{0}
\setcounter{figure}{0}
\renewcommand\thesection{\Alph{section}}
\renewcommand\theHsection{A-\thesection}
\renewcommand{\thefigure}{A.\arabic{figure}}

\section{Appendix}\label{appendix}

\subsection{Are power gains possible in basket trials when strict type-I
error control is required?}\label{sec-counterex}

Outside the research on basket trials, information borrowing has also
been studied in the context of borrowing external historical data and
other real world evidence. In that context, it is a well-known fact that
power gains cannot be achieved when requiring strict type-I error
control in the presence of a uniformly most powerful (UMP) test
\autocite{kopp-schneider_power_2020}. The ICH draft guideline on
adaptive designs states that ``{[}b{]}orrowing of external data to
inform inference requires a thorough scientific justification that
addresses the feasibility of alternative approaches not involving
borrowing (e.g., design and conduct of a fully powered trial without
using external data) and supports the relevance and quality of the
external data'' \autocite{ich_e20adaptive_2025}. As borrowing historical
data for power gains always implies type-I error inflation, this poses
the question what such a scientific justification could be.

The argument by \textcite{kopp-schneider_power_2020} applies to
situations in which a uniformly most powerful (UMP) test exists in the
class of level \(\alpha\) tests. This is the case for many important
situations, such as for endpoints from an exponential family with
one-sided hypotheses \(H_0:\theta\leq\theta_0\) versus
\(H_1:\theta>\theta_0\) and probability density function
\(f_\theta(x)\). Borrowing information from an external source means
that the external data are to be considered fixed, not random. Hence,
the power of a level \(\alpha\) test borrowing external data cannot
exceed the power of the respective UMP test.

The situation is somewhat different for basket trials. In general, no
UMP test exists in the class of level \(\alpha\) tests for
multidimensional parameters \autocite{vaart_asymptotic_1998}. In the
case of simultaneously testing the one-sided hypothesis for a vector of
parameters, this can be seen using the following counterexample: For the
sake of simplicity, consider a basket trial with \(I=2\) strata with
sample sizes \(n_1\) and \(n_2\), independent normally distributed
endpoints \(X_1\) and \(X_2\), unknown means \(\theta_1\) and
\(\theta_2\), and known variance \(\sigma^2\). Consider the global null
hypothesis \[H_0:(\theta_1,\theta_2)=(\theta_0,\theta_0)\] versus the
global alternative hypothesis

\[H_1:(\theta_1,\theta_2)\text{ with }\theta_i\geq \theta_0\text{ for all }i,\text{ and } \theta_i > \theta_0\text{ for at least one }i.\]

The conditional probability of rejecting the true null hypothesis
\(H_0\) is the FWER, the conditional probability of rejecting the true
alternative hypothesis \(H_1\) is a very general definition of the EWP.
The set of alternative hypotheses \(H_1\) is a large set which contains
the alternative hypothesis
\(H_1'\subset H_1:\theta_1=\theta^*>\theta_0,\theta_2 = \theta_0\) for
some fixed \(\theta^*\). This hypothesis is effectively a point
hypothesis in stratum 1 and ignores the information from stratum 2. The
level \(\alpha\) UMP test for these point hypotheses \(H_0\)
vs.~\(H_1'\) is simply the Z-test, so the rejection region is
\(\bar X_1>\theta_0+ z_\alpha\sigma/\sqrt n_1\) with the sample mean
\(\bar X_1\) in stratum 1 and \(z_\alpha\) being the standard normal
quantile \autocite{casella_statistical_inference_2002}. Call this test
1. On the other hand, the set \(H_1\) also contains the alternative
hypothesis \(H_1''\subset H_1:\theta_1=\theta_2=\theta^{**}>\theta_0\)
for some fixed \(\theta^{**}\). This hypothesis effectively pools the
information from the two strata and assumes they come from the same
distribution. Hence the level \(\alpha\) UMP test is again the Z test
with rejection region
\(\bar X_ {1+2}>\theta_0+ z_\alpha\sigma/\sqrt{n_1+n_2}\), where
\(\bar X_ {1+2}\) is the pooled sample mean of the whole basket trial.
Call this test 2. Now suppose that a UMP test existed for all of
\(H_1\). By definition, it would have at least the power of the test 1
and test 2 for \(H_1'\) and \(H_1''\). By the uniqueness part of the
Neyman-Pearson lemma, the rejection region of such a UMP test would be
equal to the rejection region of test 1 and equal to the rejection
region of test 2, except for a negligible set. This is a contradiction
as the two rejection regions are non-negligibly different.

This counterexample demonstrates that no UMP test exists even for
one-sided hypotheses in the context of basket trials. The reason is
clearly that the space of possible alternative hypotheses is vast, and
every test can only distribute its power to some parts of this space.
Note that the example we discussed looked at normally distributed
endpoints, whereas most basket trial methods look at binary endpoints.
For binary endpoints, it is more difficult to construct a similar
counterexample because the attainable significance levels \(\alpha\) are
dependent on the discreteness of the distribution and therefore change
when one pools the data. As the binomial distribution approximates the
normal distribution for large \(n\) and fixed \(p\), the example still
provides a hint as to the existence of a UMP test for basket trials with
binomial endpoints as well.

To some extent, this example justifies the use of information borrowing
mechanisms for obtaining power gains in basket trials. As no UMP exists
for all possible alternative hypotheses, it is sensible to define a set
of relevant alternative hypotheses and to optimize power for this set
using an optimized information borrowing approach. Nevertheless, one
should be careful as most information borrowing mechanisms are not
designed to control type-I error. If one requires strict type-I error
control, it may be better to use a multiple testing procedure, e.g.~from
the closed testing literature.

\newpage

\subsection{Further results of the comparison
study}\label{further-results-of-the-comparison-study}

\begin{figure}

\centering{

\pandocbounded{\includegraphics[keepaspectratio]{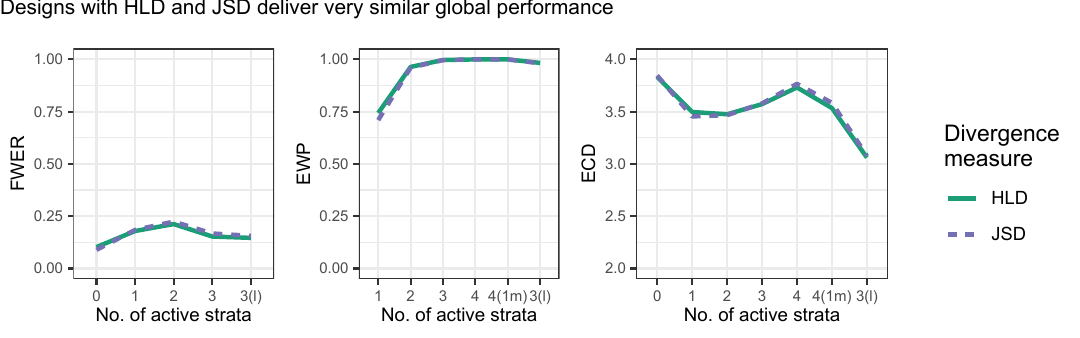}}

}

\caption{\label{fig-appendix-global-hld-vs-jsd}Global rejection rates
and correct decisions in the scenario set \texttt{sc\_bau04} comparing
the basket trial design by Fujikawa et al., which uses the
Jensen-Shannon divergence (JSD), to a modified design, which uses the
Hellinger distance (HLD) instead. Both design were optimized using the
utility function
\texttt{u\_2pow\_avg\_pen}.\newline\footnotesize \textsc{Symbols.}
4(1m): scenario with 4 active strata and `one in the middle', 3(l):
scenario with 3 active strata and `linear' increase in response, see
Table~\ref{tbl-scenario-sets}. FWER: family-wise error rate, EWP:
experiment-wise power, ECD: expected number of correct decisions.}

\end{figure}%

\begin{figure}

\centering{

\pandocbounded{\includegraphics[keepaspectratio]{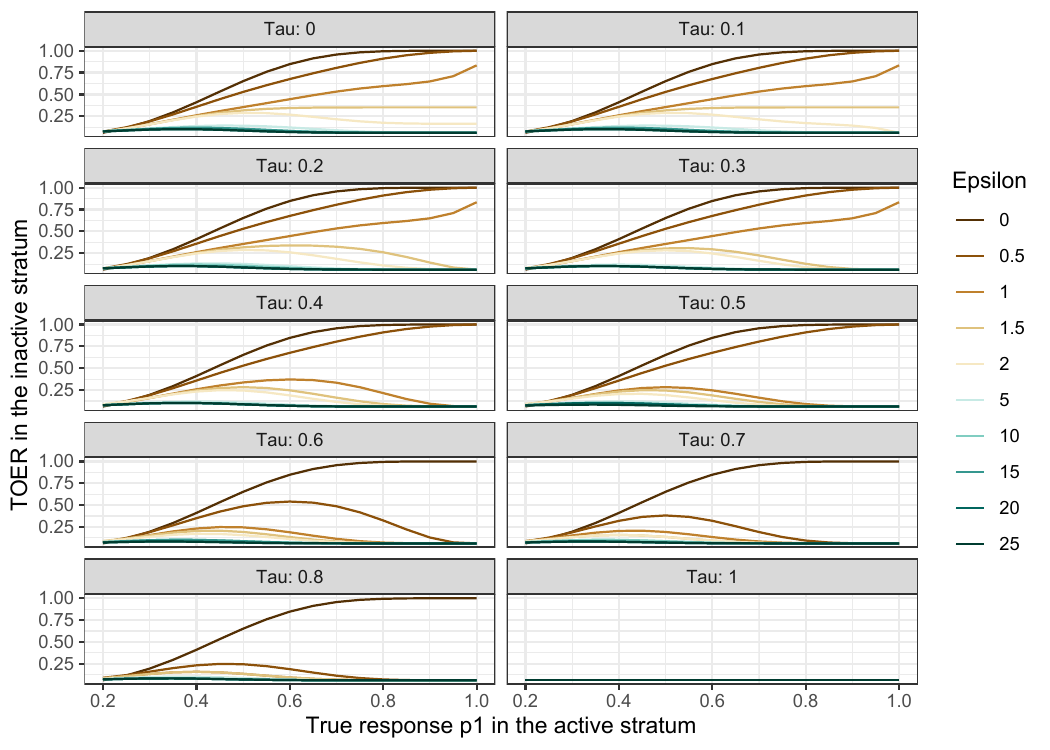}}

}

\caption{\label{fig-appendix-reject-params}Influence of borrowing
parameters on type-I error rate in a two-stratum basket trial with
per-stratum sample size \(n_i=15\) and posterior decision threshold
\(\lambda=0.95\). The inactive stratum has a response rate of
\(p_0=0.2\), the other stratum's true response rate is plotted on the
x-axis.}

\end{figure}%

\begin{figure}

\centering{

\pandocbounded{\includegraphics[keepaspectratio]{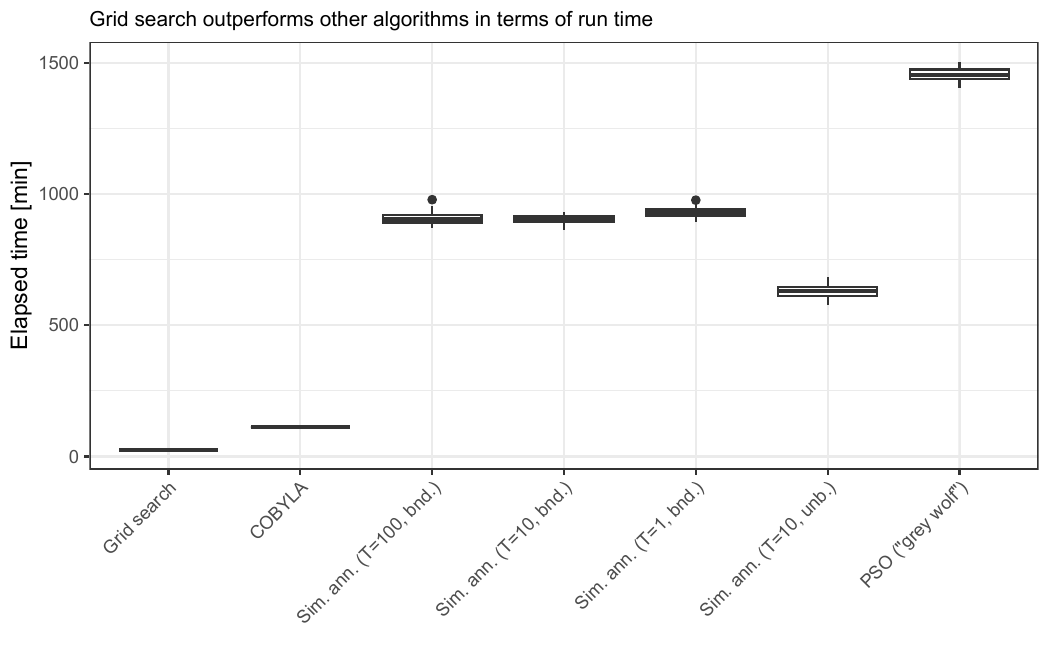}}

}

\caption{\label{fig-appendix-alg-efficiency-u-ecd-avg}Efficiency of
optimization algorithms when optimizing the utility function
\texttt{u\_ecd\_avg} in the scenario set \texttt{sc\_bau04}}

\end{figure}%

\begin{figure}

\centering{

\pandocbounded{\includegraphics[keepaspectratio]{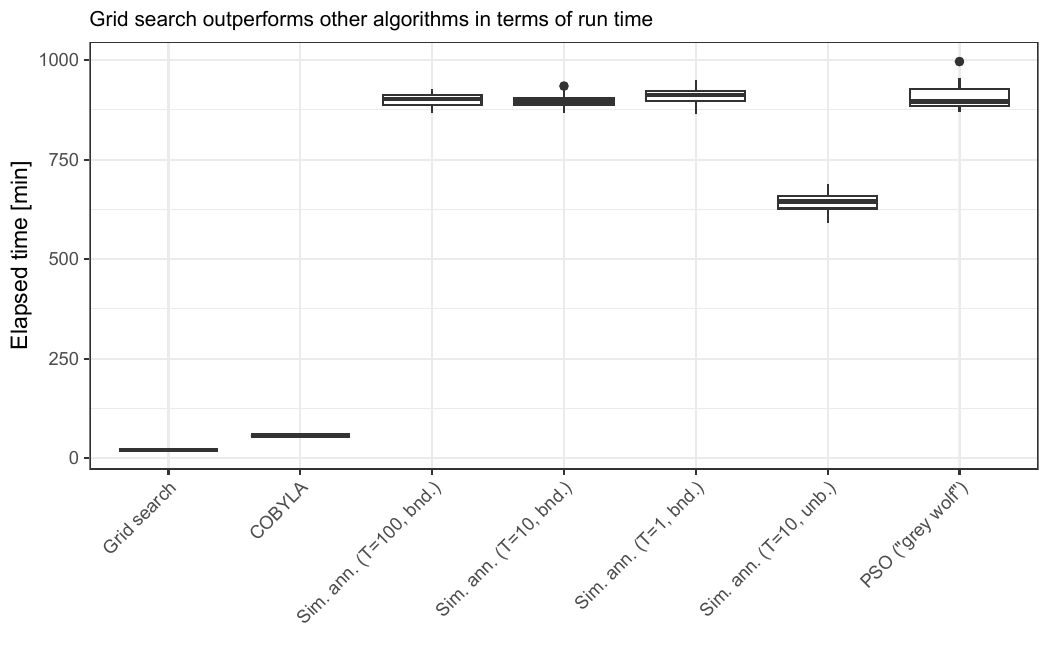}}

}

\caption{\label{fig-appendix-alg-efficiency-u-2pow-avg}Efficiency of
optimization algorithms when optimizing the utility function
\texttt{u\_2pow\_avg} in the scenario set \texttt{sc\_bau04}}

\end{figure}%

\begin{figure}

\centering{

\pandocbounded{\includegraphics[keepaspectratio]{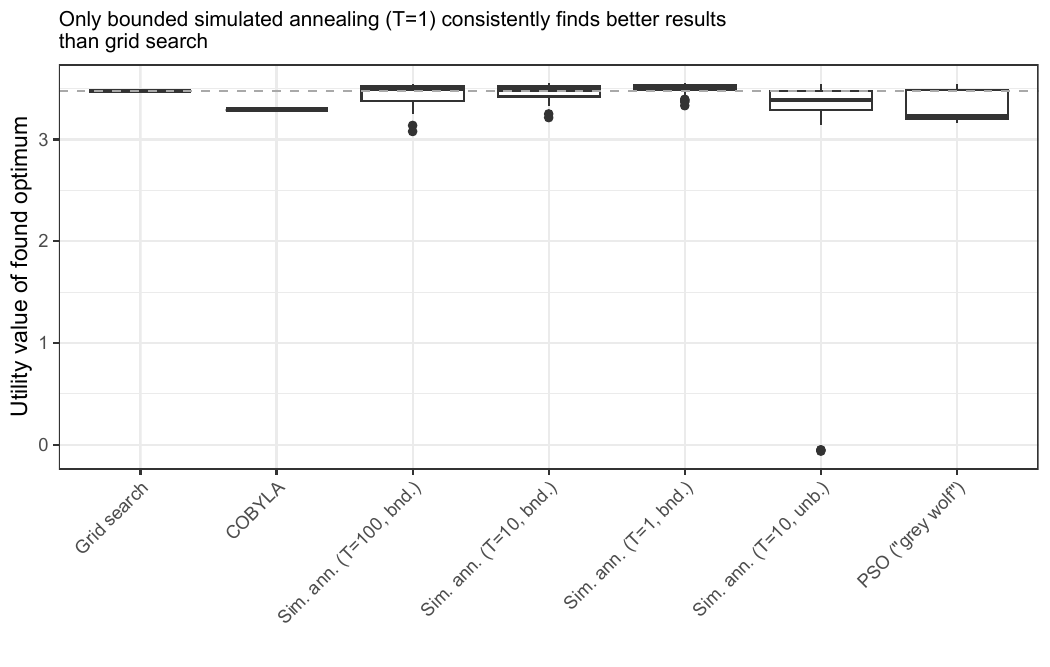}}

}

\caption{\label{fig-appendix-alg-reliability-u-ecd-avg}Reliability of
optimization algorithms in finding the optimal value of
\texttt{u\_ecd\_avg} in the scenario set \texttt{sc\_bau04}}

\end{figure}%

\begin{figure}

\centering{

\pandocbounded{\includegraphics[keepaspectratio]{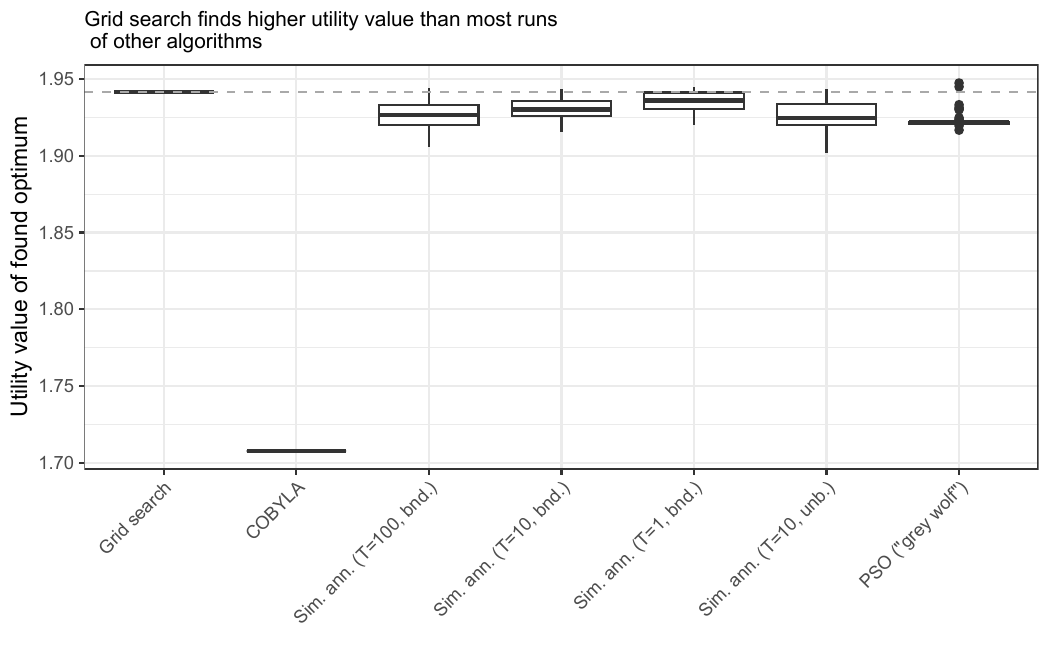}}

}

\caption{\label{fig-appendix-alg-reliability-u-2pow-avg}Reliability of
optimization algorithms in finding the optimal value of
\texttt{u\_2pow\_avg} in the scenario set \texttt{sc\_bau04}}

\end{figure}%

\begin{figure}

\centering{

\pandocbounded{\includegraphics[keepaspectratio]{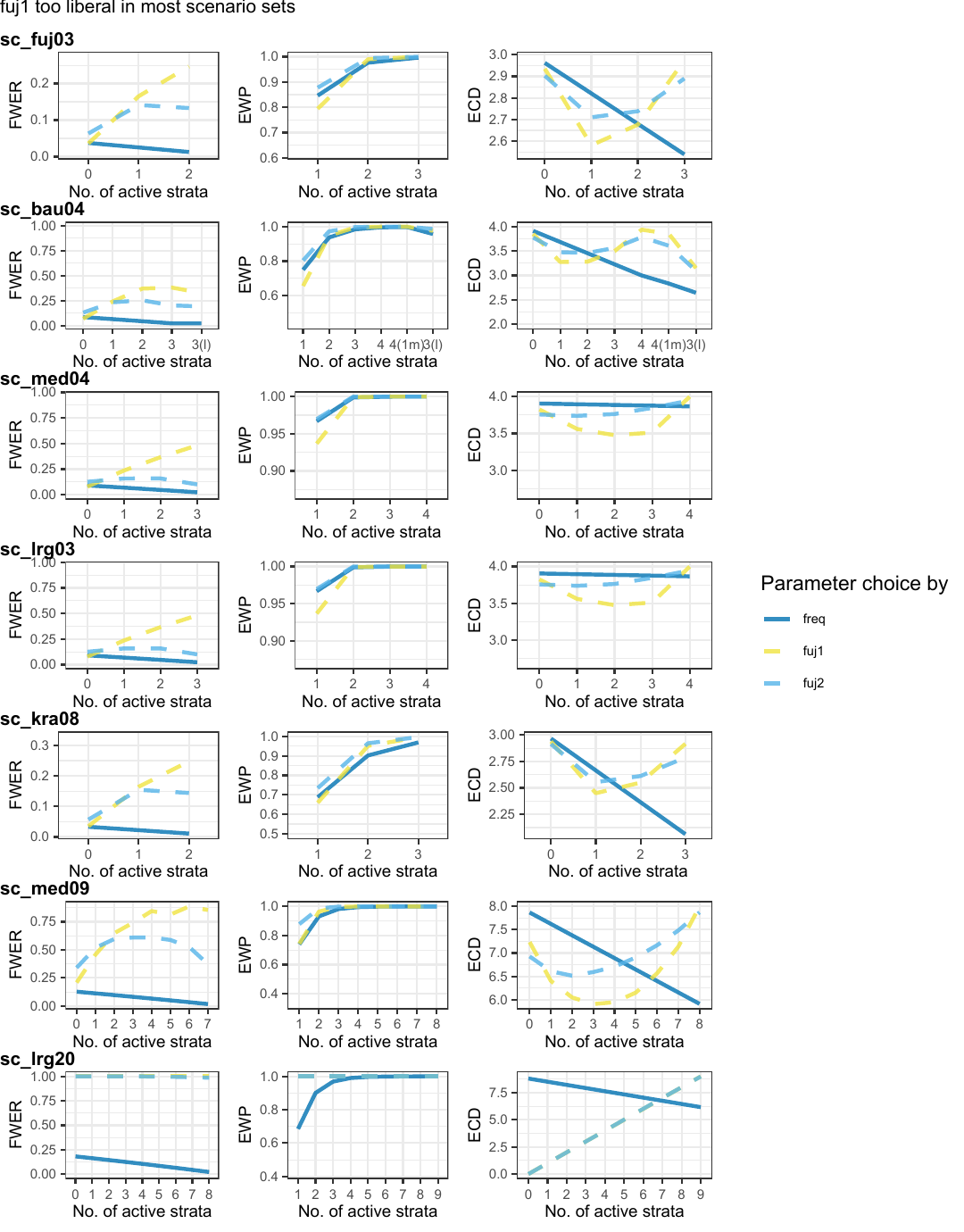}}

}

\caption{\label{fig-appendix-all-global-fuj12}Global rejection rates and
correct decisions in all scenario sets for the parameters
\texttt{fuj1/2} suggested by
\textcite{fujikawa_bayesian_2020}.\newline\footnotesize \textsc{Symbols.}
4(1m): scenario with 4 active strata and `one in the middle', 3(l):
scenario with 3 active strata and `linear' increase in response, see
Table~\ref{tbl-scenario-sets}. FWER: family-wise error rate, EWP:
experiment-wise power, ECD: expected number of correct decisions,
\texttt{freq}: frequentist basket trial design with separate unadjusted
binomial tests.}

\end{figure}%

\begin{figure}

\centering{

\pandocbounded{\includegraphics[keepaspectratio]{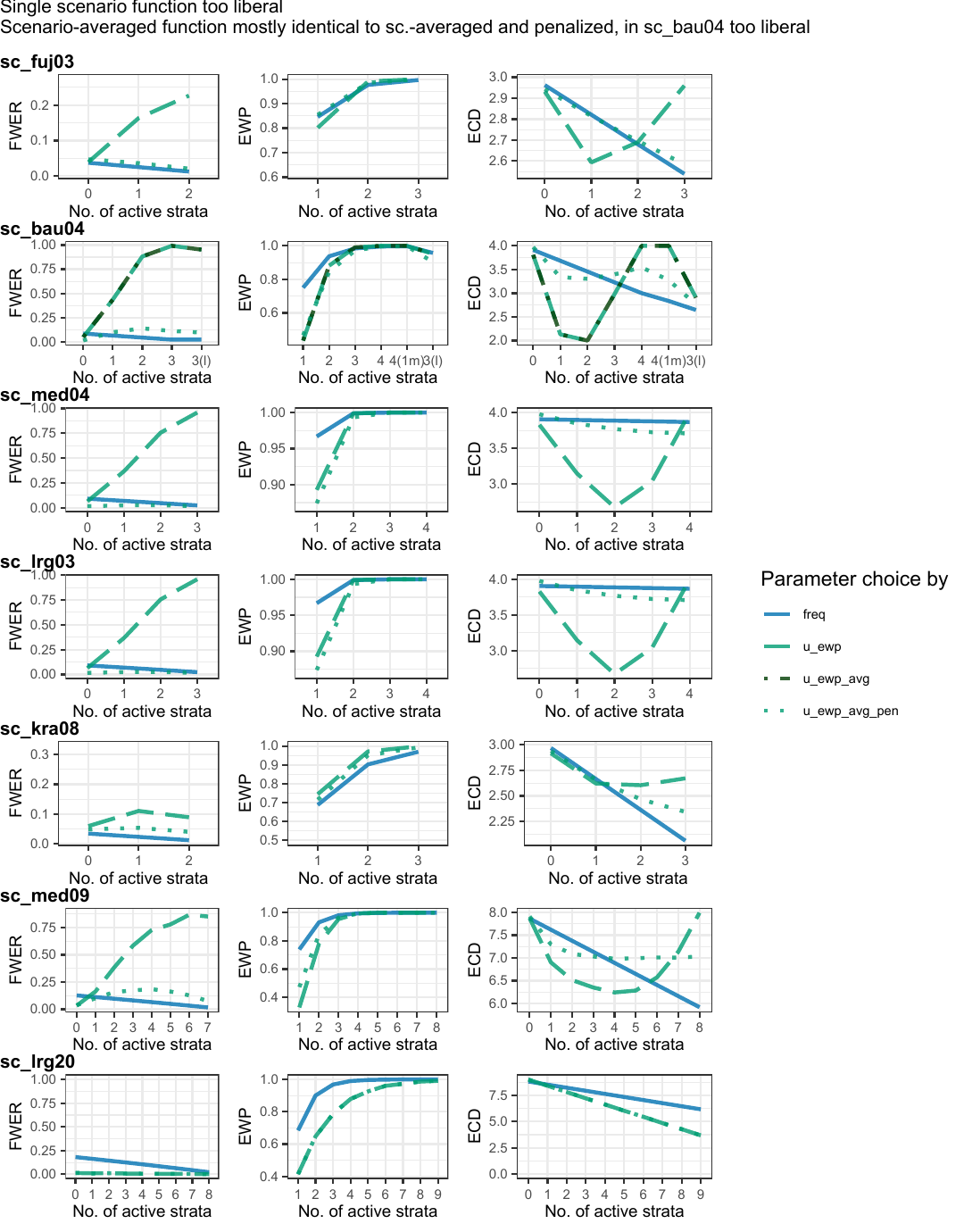}}

}

\caption{\label{fig-appendix-all-global-ewp}Global rejection rates and
correct decisions in all scenario sets for the parameters suggested by
the \texttt{u\_ewp\_*} family. In all scenario sets except
\texttt{sc\_bau04}, the functions \texttt{u\_ecd\_avg} and
\texttt{u\_ecd\_avg\_pen} found the same parameters. In
\texttt{sc\_bau04}, \texttt{u\_ecd} and \texttt{u\_ecd\_avg} found the
same parameters.\newline\footnotesize \textsc{Symbols.} 4(1m): scenario
with 4 active strata and `one in the middle', 3(l): scenario with 3
active strata and `linear' increase in response, see
Table~\ref{tbl-scenario-sets}. FWER: family-wise error rate, EWP:
experiment-wise power, ECD: expected number of correct decisions,
\texttt{freq}: frequentist basket trial design with separate unadjusted
binomial tests, \texttt{u\_*}: utility functions as defined in
Section~\ref{sec-utility-functions}.}

\end{figure}%

\begin{figure}

\centering{

\pandocbounded{\includegraphics[keepaspectratio]{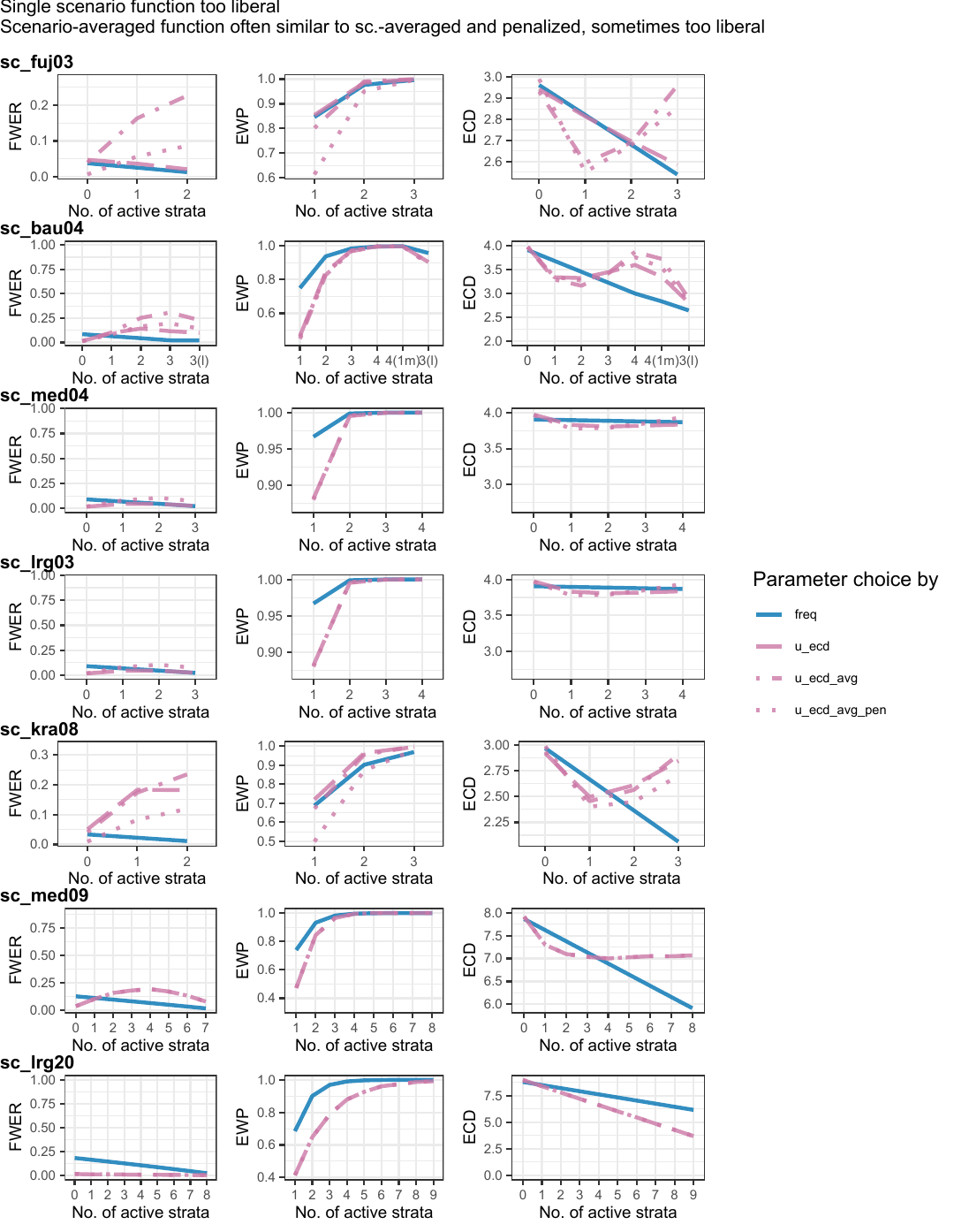}}

}

\caption{\label{fig-appendix-all-global-ecd}Global rejection rates and
correct decisions in all scenario sets for the parameters suggested by
the \texttt{u\_ecd\_*} family. In \texttt{sc\_med04},
\texttt{sc\_lrg03}, \texttt{sc\_med09} and \texttt{sc\_lrg20}, the
functions \texttt{u\_ecd\_avg} and \texttt{u\_ecd\_avg\_pen} found the
same parameters.\newline\footnotesize \textsc{Symbols.} 4(1m): scenario
with 4 active strata and `one in the middle', 3(l): scenario with 3
active strata and `linear' increase in response, see
Table~\ref{tbl-scenario-sets}. FWER: family-wise error rate, EWP:
experiment-wise power, ECD: expected number of correct decisions,
\texttt{freq}: frequentist basket trial design with separate unadjusted
binomial tests, \texttt{u\_*}: utility functions as defined in
Section~\ref{sec-utility-functions}.}

\end{figure}%

\begin{figure}

\centering{

\pandocbounded{\includegraphics[keepaspectratio]{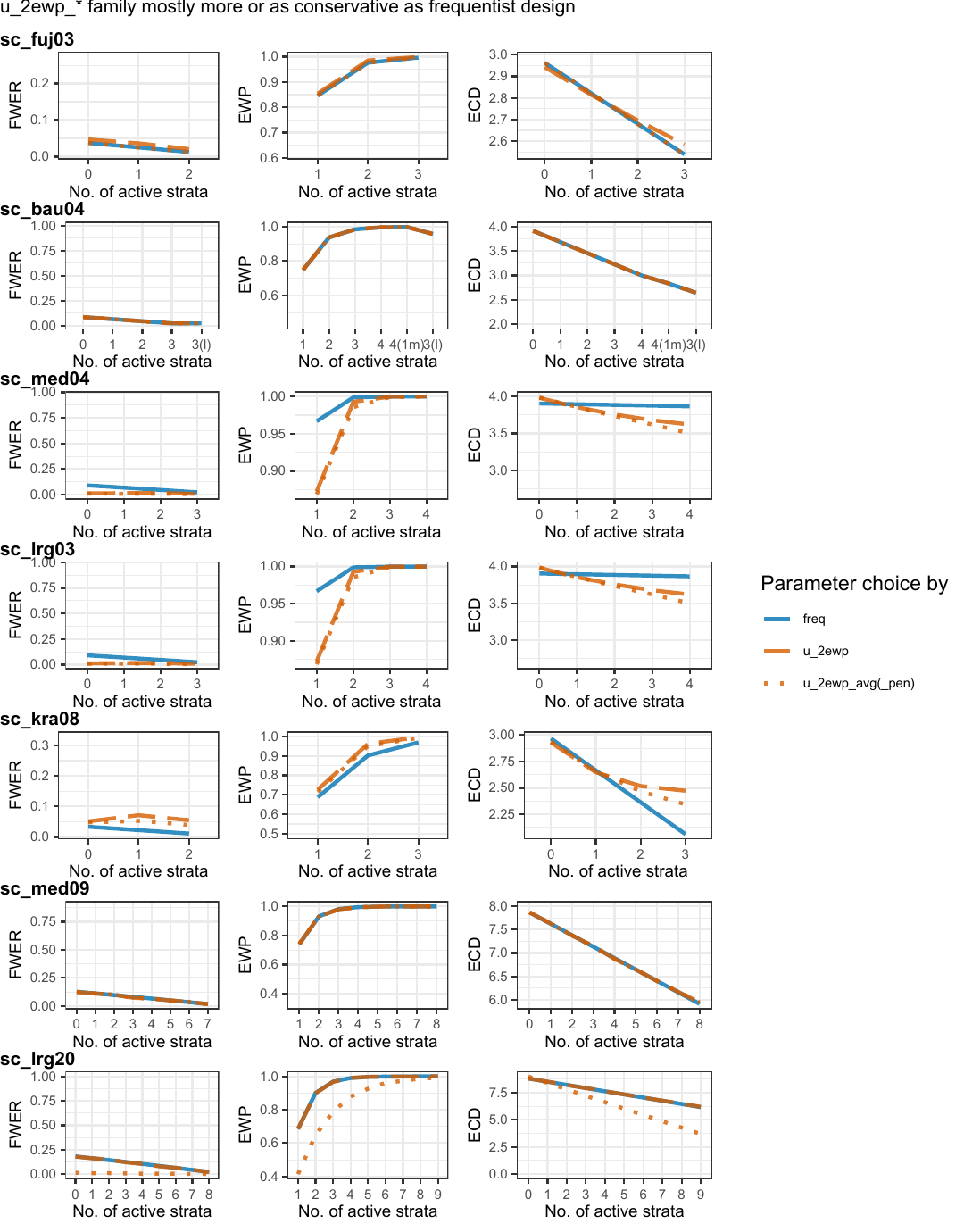}}

}

\caption{\label{fig-appendix-all-global-2ewp}Global rejection rates and
correct decisions in all scenario sets for the parameters found by the
\texttt{u\_2ewp\_*} family. In all scenario sets, the functions
\texttt{u\_2ewp\_avg} and \texttt{u\_2ewp\_avg\_pen} found the same
parameters.\newline\footnotesize \textsc{Symbols.} 4(1m): scenario with
4 active strata and `one in the middle', 3(l): scenario with 3 active
strata and `linear' increase in response, see
Table~\ref{tbl-scenario-sets}. FWER: family-wise error rate, EWP:
experiment-wise power, ECD: expected number of correct decisions,
\texttt{freq}: frequentist basket trial design with separate unadjusted
binomial tests, \texttt{u\_*}: utility functions as defined in
Section~\ref{sec-utility-functions}.}

\end{figure}%

\begin{figure}

\centering{

\pandocbounded{\includegraphics[keepaspectratio]{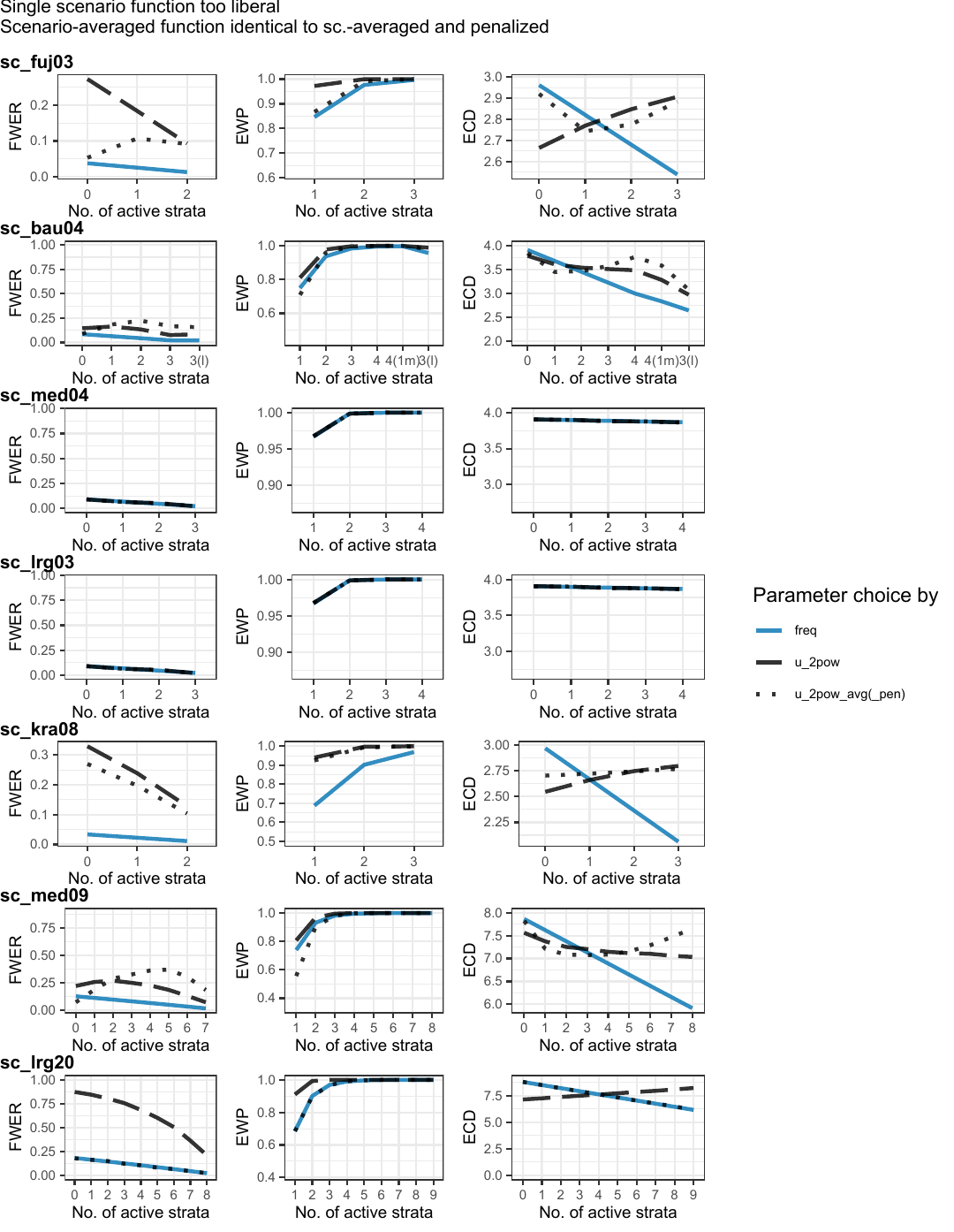}}

}

\caption{\label{fig-appendix-all-global-2pow}Global rejection rates and
correct decisions in all scenario sets for the parameters suggested by
the \texttt{u\_2pow\_*} family. In all scenario sets, the functions
\texttt{u\_2pow\_avg} and \texttt{u\_2pow\_avg\_pen} found the same
parameters.\newline\footnotesize \textsc{Symbols.} 4(1m): scenario with
4 active strata and `one in the middle', 3(l): scenario with 3 active
strata and `linear' increase in response, see
Table~\ref{tbl-scenario-sets}. FWER: family-wise error rate, EWP:
experiment-wise power, ECD: expected number of correct decisions,
\texttt{freq}: frequentist basket trial design with separate unadjusted
binomial tests, \texttt{u\_*}: utility functions as defined in
Section~\ref{sec-utility-functions}.}

\end{figure}%



\end{document}